\documentclass[twocolumn,preprint]{aastex631}
\usepackage{natbib, twoopt}
\usepackage{float}
\usepackage{graphicx}
\usepackage{epstopdf}
\usepackage{enumitem}
\usepackage[caption=false]{subfig}
\usepackage{xcolor}
\newcommand{\revfont}{
  \bfseries
  \color{red}
}
\DeclareTextFontCommand{\rev}{\revfont}
\graphicspath{{Figures/}}
\RequirePackage{import}

\begin{document}

\title{Extreme Red-wing Enhancements of UV Lines During the 2022 March 30 X1.3 Solar Flare}

    \author{Yan Xu}
	\affil{Institute for Space Weather Sciences, New Jersey Institute of Technology, 323 Martin Luther King Boulevard, Newark, NJ 07102-1982}
	\affil{Big Bear Solar Observatory, New Jersey Institute of Technology, 40386 North Shore Lane, Big Bear City, CA 92314-9672, USA}
	
    \author{Graham~S. Kerr}
	\affil{NASA Goddard Space Flight Center, Heliophysics Science Division, Code 671, 8800 Greenbelt Rd., Greenbelt, MD 20771, USA}
 	\affil{Department of Physics, Catholic University of America, 620 Michigan Avenue, Northeast, Washington, DC 20064, USA}

    \author{Vanessa Polito}
	\affiliation{Bay Area Environmental Research Institute, NASA Research Park,  Moffett Field, CA 94035-0001, USA}
    \affiliation{Lockheed Martin Solar and Astrophysics Laboratory, Building 252, 3251 Hanover Street, Palo Alto, CA 94304, USA}

    \author{Nengyi Huang}
	\affil{Institute for Space Weather Sciences, New Jersey Institute of Technology, 323 Martin Luther King Boulevard, Newark, NJ 07102-1982}
	  \affil{Big Bear Solar Observatory, New Jersey Institute of Technology, 40386 North Shore Lane, Big Bear City, CA 92314-9672, USA}

	\author{Ju Jing}
	\affil{Institute for Space Weather Sciences, New Jersey Institute of Technology, 323 Martin Luther King Boulevard, Newark, NJ 07102-1982}
	\affil{Big Bear Solar Observatory, New Jersey Institute of Technology, 40386 North Shore Lane, Big Bear City, CA 92314-9672, USA}
	
	\author{Haimin Wang}
	\affil{Institute for Space Weather Sciences, New Jersey Institute of Technology, 323 Martin Luther King Boulevard, Newark, NJ 07102-1982}
	\affil{Big Bear Solar Observatory, New Jersey Institute of Technology, 40386 North Shore Lane, Big Bear City, CA 92314-9672, USA}

\begin{abstract}
Here we present the study of a compact emission source during an X1.3 flare on 2022-March-30. Within a $\sim41$~s period (17:34:48 UT to 17:35:29 UT), IRIS observations show spectral lines of \ion{Mg}{2}, \ion{C}{2} and \ion{Si}{4} with extremely broadened, asymmetric red-wings. This source of interest (SOI) is compact, $\sim$ 1\arcsec.6, and is located in the wake of a passing ribbon. Two methods were applied to measure the Doppler velocities associated with these red wings: spectral moments and multi-Gaussian fits. The spectral moments method considers the averaged shift of the lines, which are 85 km s$^{-1}$, 125 km s$^{-1}$ and 115 km s$^{-1}$ for the \ion{Mg}{2}, \ion{C}{2} and \ion{Si}{4} lines respectively. The red-most Gaussian fit suggests a Doppler velocity up to $\sim$160 km s$^{-1}$ in all of the three lines. Downward mass motions with such high speeds are very atypical, with most chromospheric downflows in flares on the order 10-100 km s$^{-1}$. Furthermore, EUV emission is strong within flaring loops connecting two flare ribbons located mainly to the east of the central flare region. The EUV loops that connect the SOI and its counterpart source in the opposite field are much less brightened, indicating that the density and/or temperature is comparatively low. These observations suggest a very fast downflowing plasma in transition region and upper chromosphere, that decelerates rapidly since there is no equivalently strong shift of the \ion{O}{1} chromospheric lines. This unusual observation presents a challenge that models of the solar atmosphere's response to flares must be able to explain. \\
\end{abstract}

\keywords{Sun: activity --- Sun: flares --- Sun: chromosphere}

\newpage

\section{Introduction}
\label{Sect:intro}
During a flare, lower layers of solar atmosphere, such as the chromosphere and/or transition region are heated rapidly by non-thermal particles \citep{Carmichael1964, Sturrock1968, Hirayama1974, Kopp1976, Hudson1972, Emslie1978, Holman2011, Kontar2011}, by thermal conduction from a flare-heated corona \citep{Antiochos1978,Cheng1983, MacNeice1986}, or by Alfv\'en waves \citep{Fletcher2008,2016ApJ...818L..20R,2018ApJ...853..101R, Kerr2016}. This leads to strong temperature gradients in these layers, when the radiative losses are smaller than the heating rate. As a consequence, the heated atmosphere expands both upward and downward, known as chromospheric evaporations and condensations, respectively. Chromospheric evaporation drives chromospheric material into the corona, filling the  flare loops and increasing their emission measure such that they subsequently brighten.

\begin{figure*}[th]
\center
\vspace{-2mm}
\includegraphics[width=\textwidth]{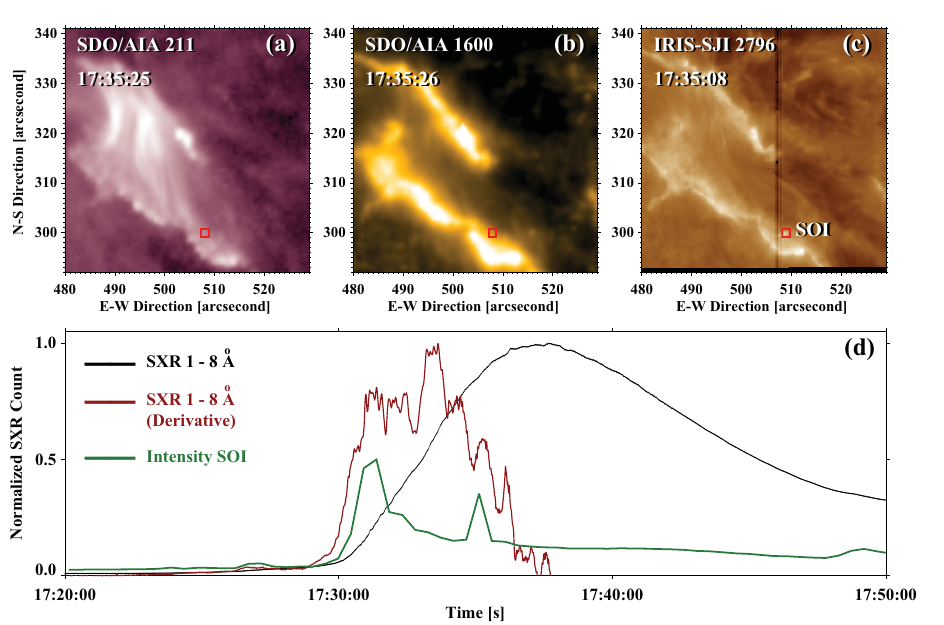}
\caption{\textsl{Panel (a): SDO/AIA 211~\AA\ image taken at 17:35:25 UT, showing two major flare ribbons and flaring loops. Panel (b): SDO/AIA 1600~\AA\ image taken at 17:35:26 UT. Panel (c): IRIS SJI 2796~\AA\ image in the same FOV as the AIA image, taken at 17:35:08 UT. The dark line indicates the slit position and the small red box SOI indicates the region of interest. Panel (d): Light curves of GOES soft X-ray (SXR, in black), derivative of SXR (red), UV-2796 in SOI (green).}}
\label{Fig:overview}
\end{figure*}

The upward flow of chromospheric material (chromospheric evaporation) is usually observed in spectral lines as blueshifted emission. Evaporation has been observed in many flares, with coronal lines exhibiting Doppler shifts in excess of $100$ km~s$^{-1}$ \citep[see for example this review of extreme-UV (EUV) flare emission:][]{2015SoPh..290.3399M}. The upward velocity has been observed to reach as high as 300 km s$^{-1}$ in the \ion{Fe}{19} line \citep{Teriaca2003, Milligan2006a, Milligan2006b, Teriaca2006} and 400 km s$^{-1}$ in the \ion{Ca}{19} line \citep{Antonucci1982, Antonucci1983, Wuelser1994, Ding1996a, Doschek2005}. Using observations of \ion{Fe}{21} 1354.1~\AA\ line ($T\sim11$~MK), recent studies such as \citet{Battaglia2015, Graham2015, Tian2014, Tian2015, Tian2018, Polito2015, Polito2016} found blueshifted emission up to 350 km s$^{-1}$. Compared to spectral analyses, observations of chromospheric evaporation obtained from imaging are rare. Upward motions (up to 500 km s$^{-1}$) of X-ray sources have been detected using the Reuven Ramaty High Energy Solar Spectroscopic Imager  \citep[RHESSI;][]{RHESSI} observations \citep{Liu2006, Ning2009} and using \textsl{Hinode}/X-Ray Telescope observations \citep{Nitta2012, Zhang2013}. Chromospheric evaporation can be characterized as an `explosive' event if supersonic up flows are present, or as a `gentle' event if only subsonic upflows are present. Hydrodynamic modelling suggests that one of the determinant is the energy flux carried by energetic electrons \citep[e.g.][]{Fisher1985a, Fisher1985b, Fisher1985c, Fisher1989}. A threshold of $\sim10^{10}$~erg~cm$^{-2}$~s$^{-1}$ is sometimes considered to separate the `explosive' and `gentle' regimes. The low-energy cutoff of the non-thermal particle distribution is also known to be important \citep{2015ApJ...808..177R,Polito2018}.

As a signature of chromospheric condensation, H$\alpha$ red-wing asymmetries were first reported by \citet{Waldmeier1941} and \citet{Ellison1943}. Those observations were interpreted as representations of downflows with velocities generally between 10 to 100 km s$^{-1}$ \citep{Ichimoto1984, Fisher1985a}. Later, the correlation between H$\alpha$ asymmetry and flare heating was confirmed by \citet{Wuelser1987}, \citet{Wuelser1989} and \citet{Canfield1987}. High spatial, spectral, and temporal Observations of redshifted spectral lines, or lines with red-wing asymmetries, from the Interface Region Imaging Spectrograph \citep[IRIS;][]{IRIS} have suggested typical downflow speeds on the order of $\sim10-100$ km s$^{-1}$, in C- to X-class flares, obtained from resonance lines of \ion{Si}{4} and \ion{Mg}{2}, or from numerous other chromospheric transitions \citep{Kerr2015, Liu2015, Graham2015, Graham2020, Lorinvcik2022,
Yu2020, Lorincik2022,2023arXiv230811275W}. See also the reviews by \cite{2021SoPh..296...84D} and \cite{2022FrASS...960856K}, and references therein. Downflows can exhibit as either full line shifts, red wing asymmetries, or as separate redsifted components. In addition to downflows located at flare footpoints, larger magnitude downflows (as high as 180 km s$^{-1}$) have been observed in flaring loops or in coronal rain, where the density is relatively lower \citep{Kleint2014, Lacatus2017}. \citet{Tian2015} reported downflows of up to 60 km s$^{-1}$ seen in \ion{Si}{4} line, but with some 150 km s$^{-1}$ blobs. For short duration redshifts, the \ion{Si}{4} lines represent both draining of loops as they cool, and chromospheric condensations in footpoints \citep{Tian2018}.

Mass flows in flares have also been studied extensively via numerical modelling, primarily via field-aligned (1D) loop modelling. Such models include energy transport by both electron beams and via thermal conduction (some of which also include non-local radiative heating and cooling, sometimes referred to as `backwarming', via plane-parallel radiation transport) and have been largely successful in reproducing the typical magnitudes of mass flows (that is, up to several hundred 100~km~s$^{-1}$ upflows, and downflows in the region 10-100~km~s$^{-1}$), but have been less than successful in reproducing the timescales of mass flows (which are inferred from lifetimes of Doppler shifts and asymmetries). Evaporation poses a larger problem for models, with upflows rapidly subsiding soon after the cessation of flare energy injection, whereas observed blueshifts persist for up to 10 minutes in single IRIS pixels \citep[e.g.][]{Graham2015}. Chromospheric condensations also subside somewhat too fast in models compared to observations, but the problem is not as stark as the evaporation timescales \cite[e.g.][]{Graham2020}. Discrepancies between timescales of upflows and downflows in observations and models are somewhat mitigated by multi-thread loop models \citep{Reep2018b}. Models have suggested that the downflowing material is located within a dense plug of material originating in the transition region that subsequently decelerates as it accrues more mass, such that it is propagating with a speed on the order of up to few 10s km~s$^{-1}$. If studied using narrow lines then this can appear as a wholly separate component adjacent to a `stationary component' \cite[e.g.][]{Graham2020}. For a comprehensive discussion see the review by \cite{2022FrASS...960856K}.

In this study, we present IRIS observations of a very localised source of enhanced emission, which re-brightens a few minutes after its initial brightening. Spectral lines originating from this source exhibit extremely large red wings, from which we infer strong redshifts in the transition region and upper chromosphere. These atypical, very transient, redshifts suggest downflowing material with a speed exceeding 100~km~s$^{-1}$, presenting a challenge to standard flare models.

\begin{figure*}
	\centering
	\vbox{
	\hbox{
	\subfloat{\includegraphics[width = 0.315\textwidth, clip = true, trim = 0.cm 0.cm 0.cm 0.cm]{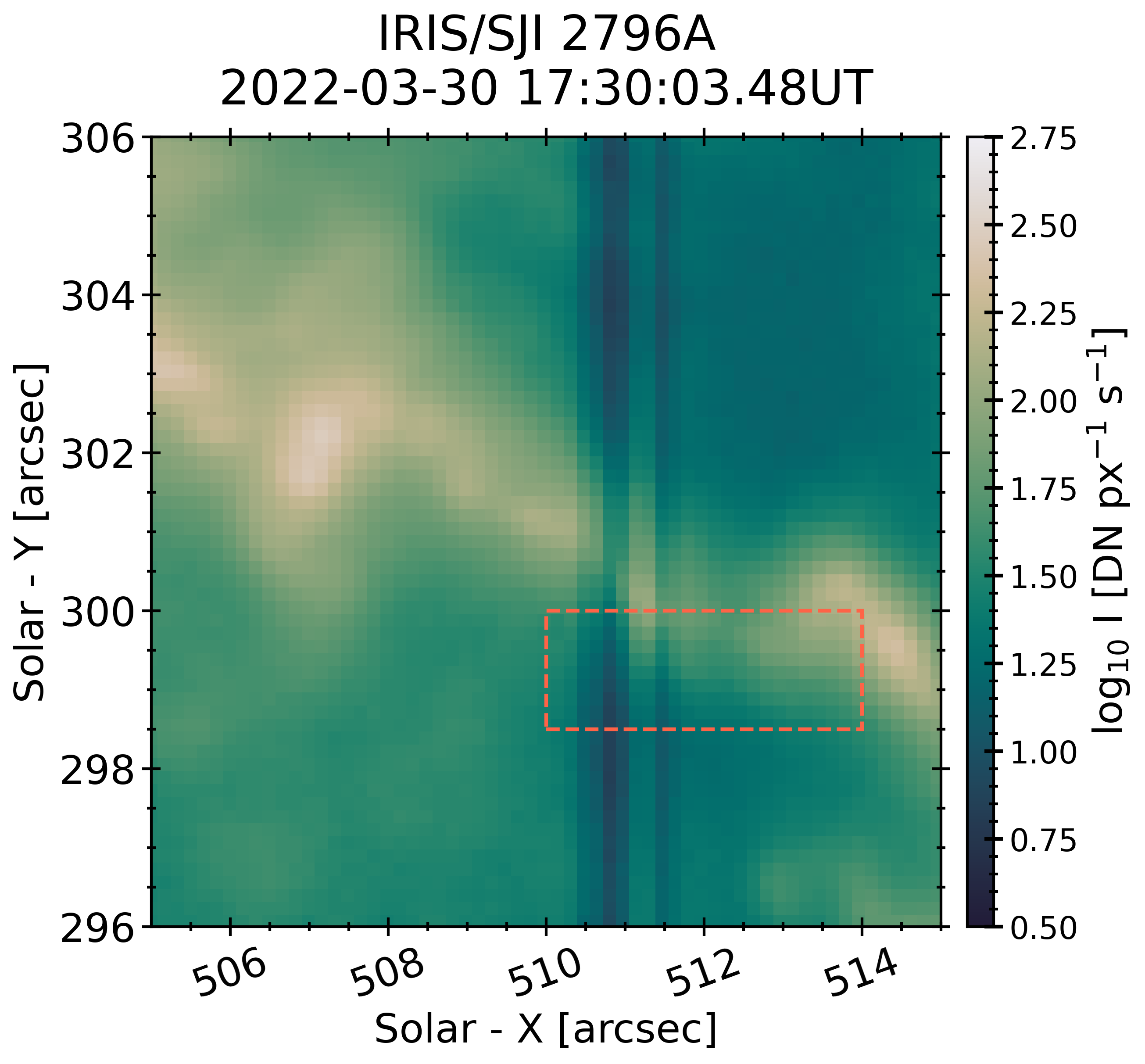}}
	\subfloat{\includegraphics[width = 0.315\textwidth, clip = true, trim = 0.cm 0.cm 0.cm 0.cm]{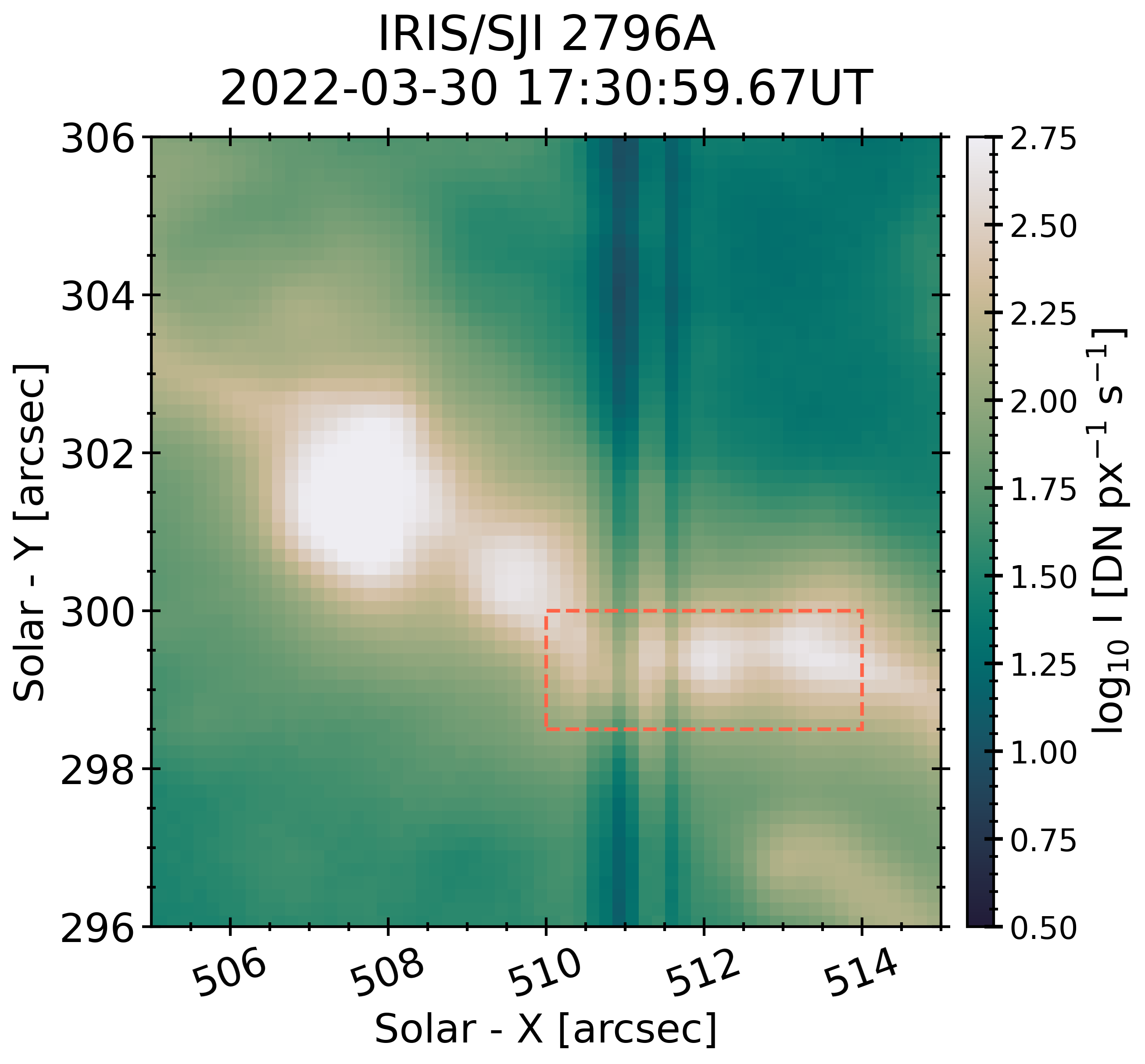}}
 \subfloat{\includegraphics[width = 0.315\textwidth, clip = true, trim = 0.cm 0.cm 0.cm 0.cm]{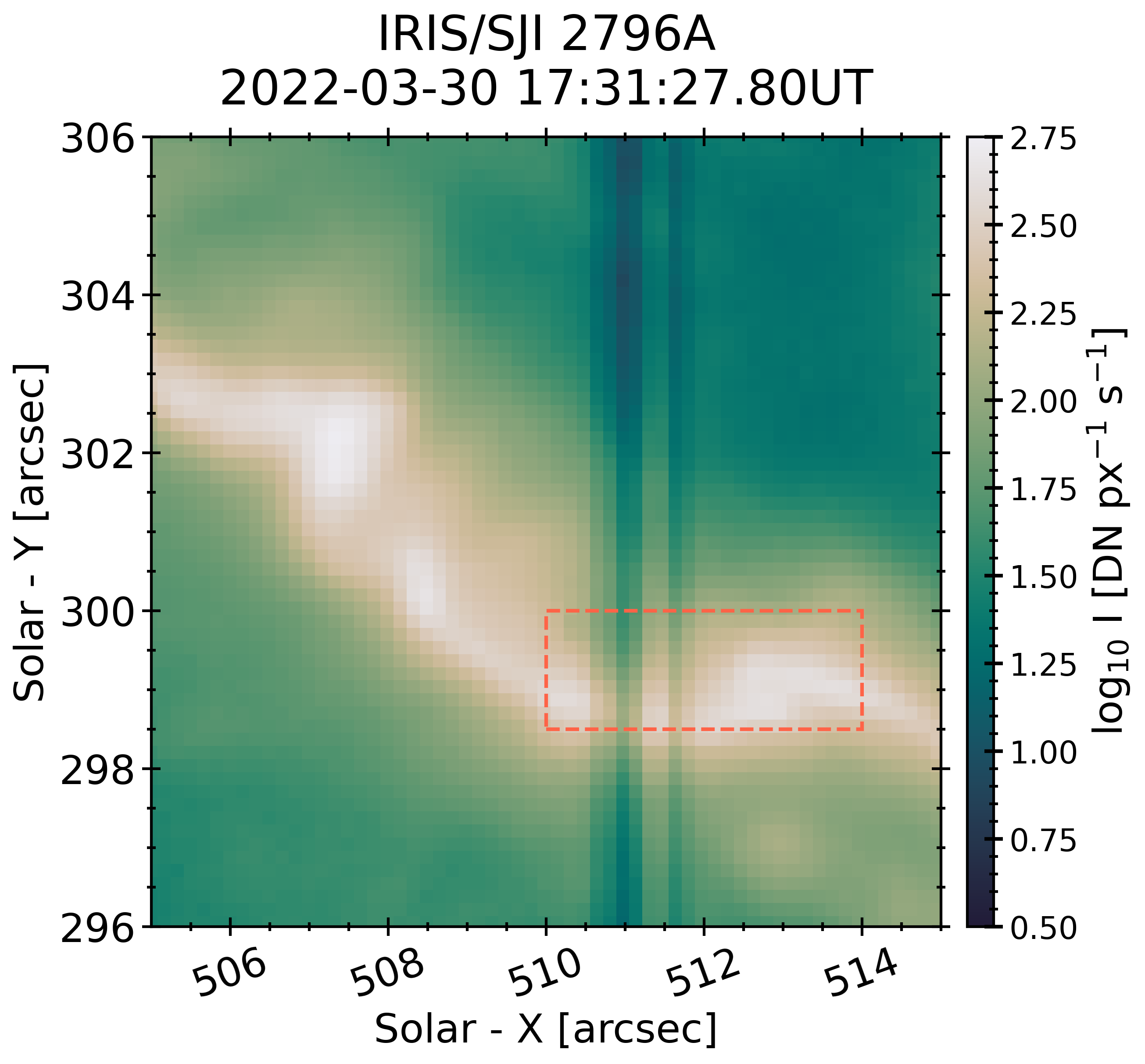}}
        }
        }
        \vbox{
	\hbox{
	\subfloat{\includegraphics[width = 0.315\textwidth, clip = true, trim = 0.cm 0.cm 0.cm 0.cm]{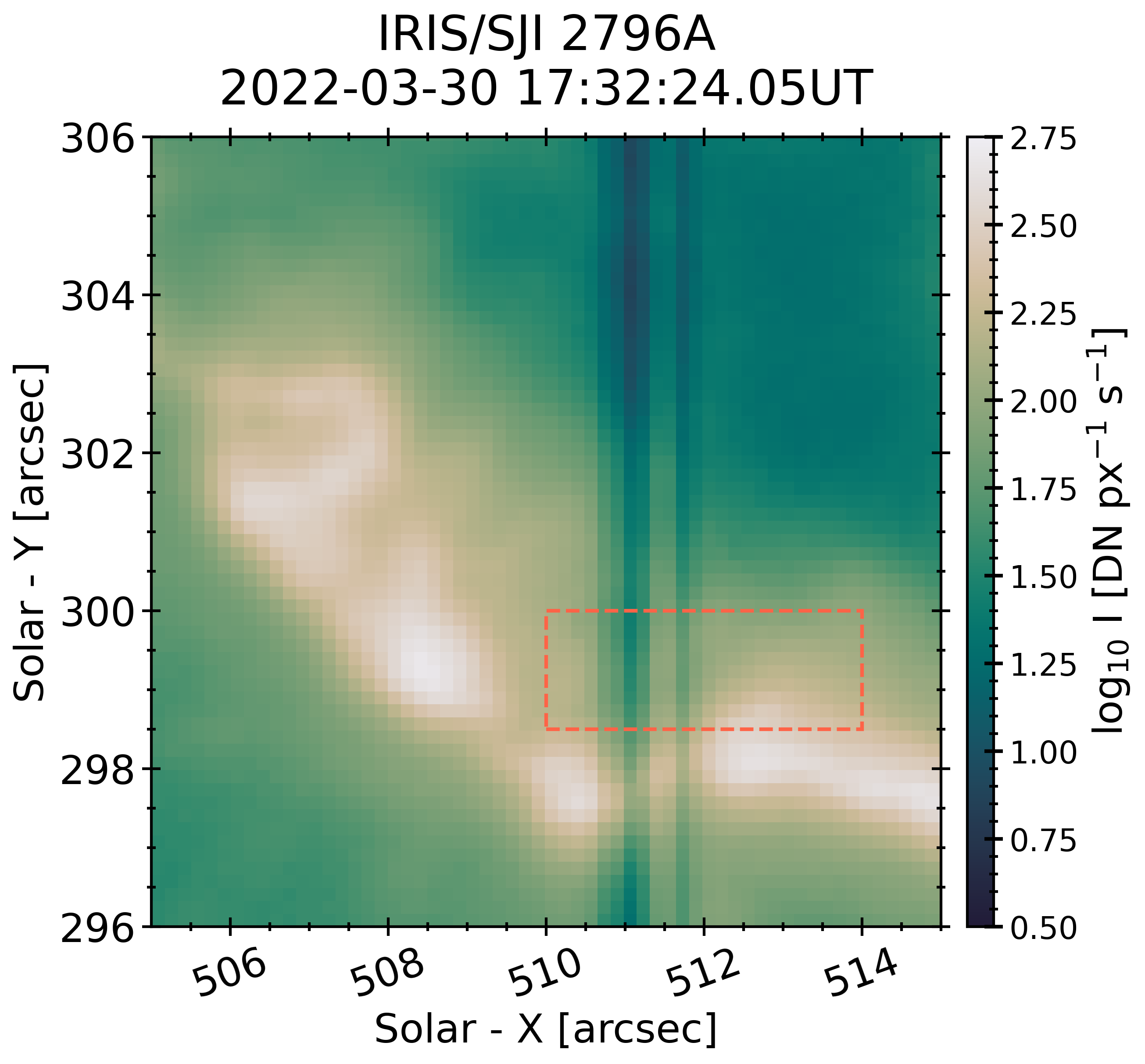}}
	\subfloat{\includegraphics[width = 0.315\textwidth, clip = true, trim = 0.cm 0.cm 0.cm 0.cm]{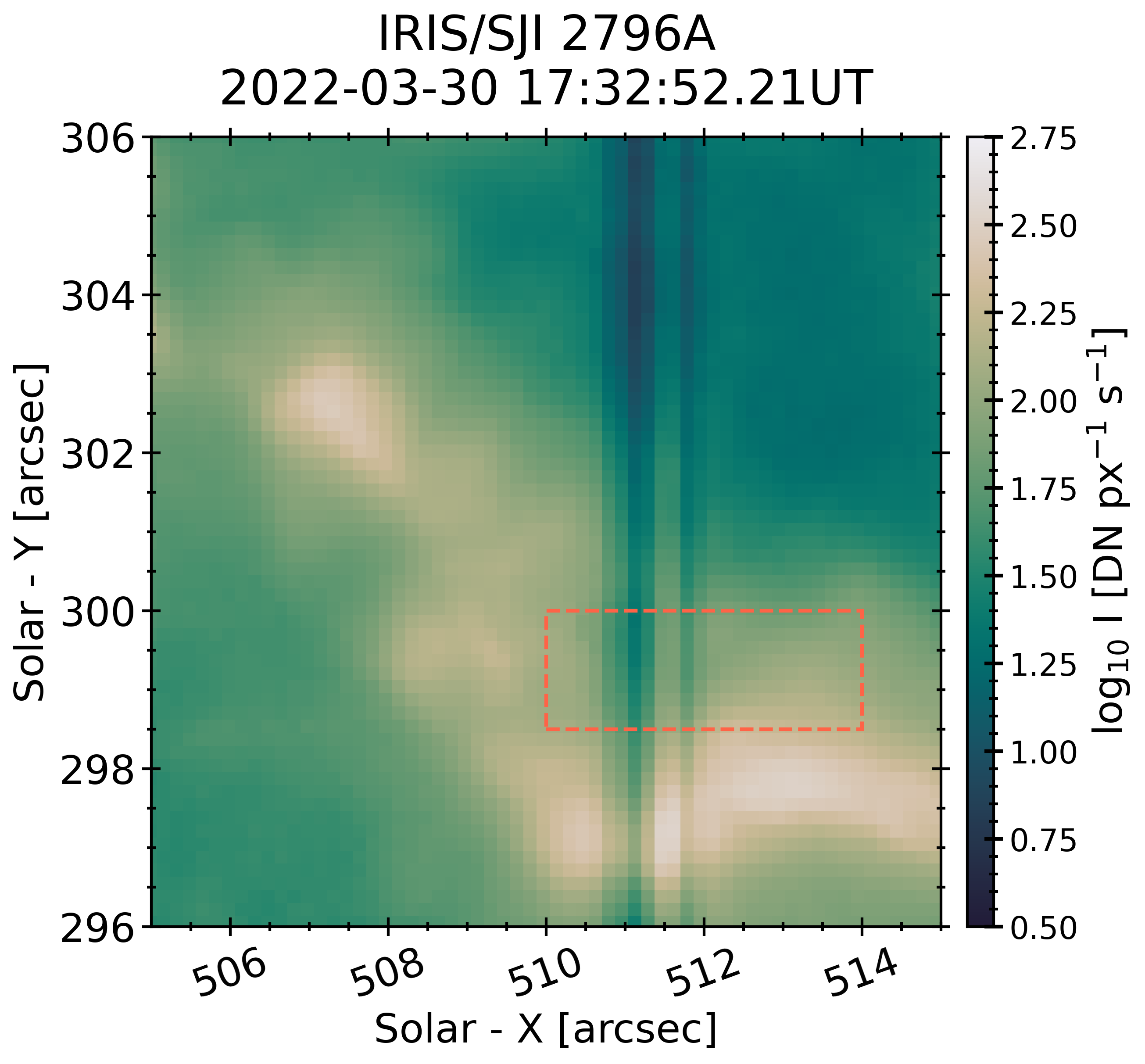}}
 \subfloat{\includegraphics[width = 0.315\textwidth, clip = true, trim = 0.cm 0.cm 0.cm 0.cm]{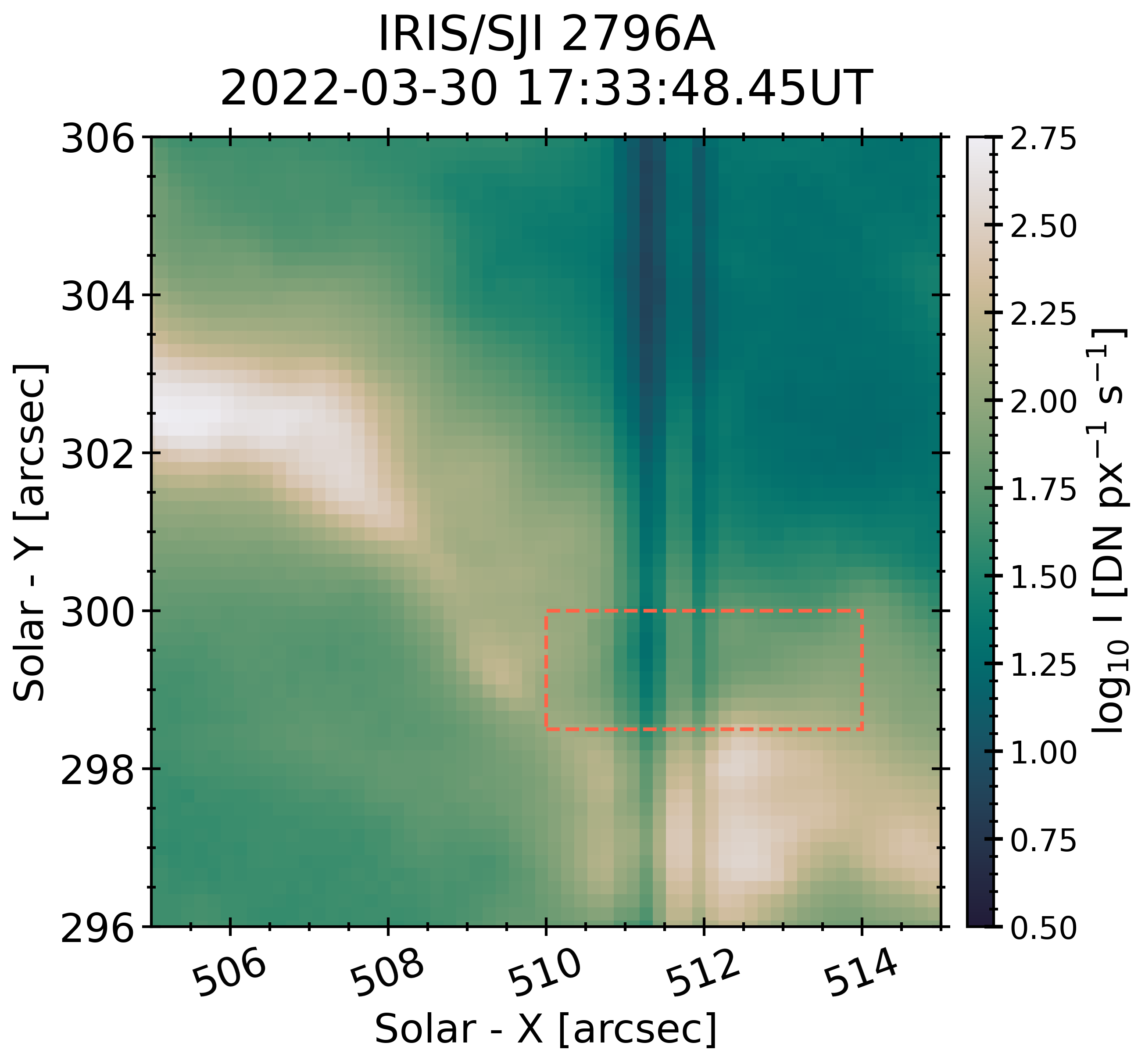}}
        }
        }
        \vbox{
	\hbox{
	\subfloat{\includegraphics[width = 0.315\textwidth, clip = true, trim = 0.cm 0.cm 0.cm 0.cm]{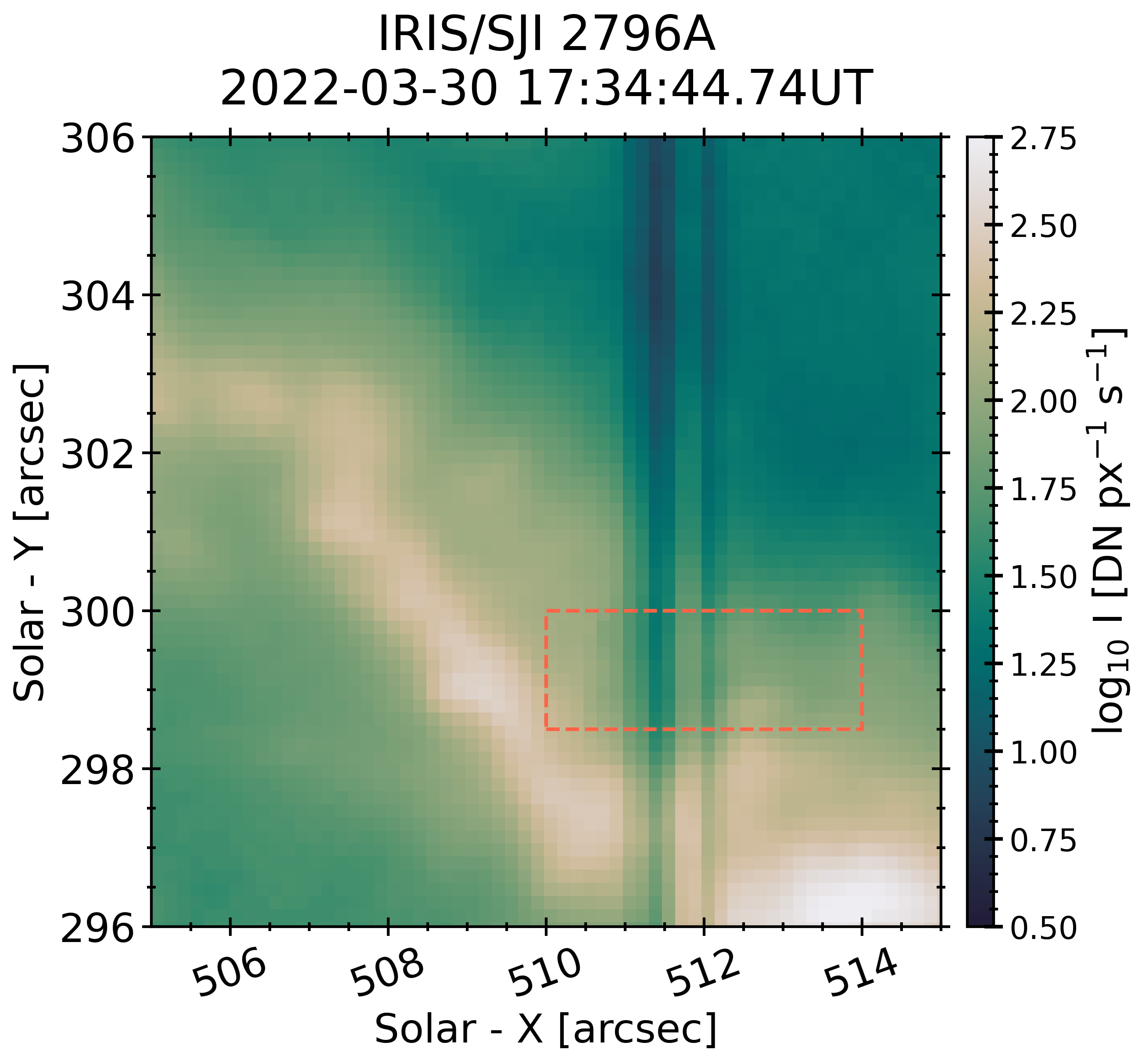}}
	\subfloat{\includegraphics[width = 0.315\textwidth, clip = true, trim = 0.cm 0.cm 0.cm 0.cm]{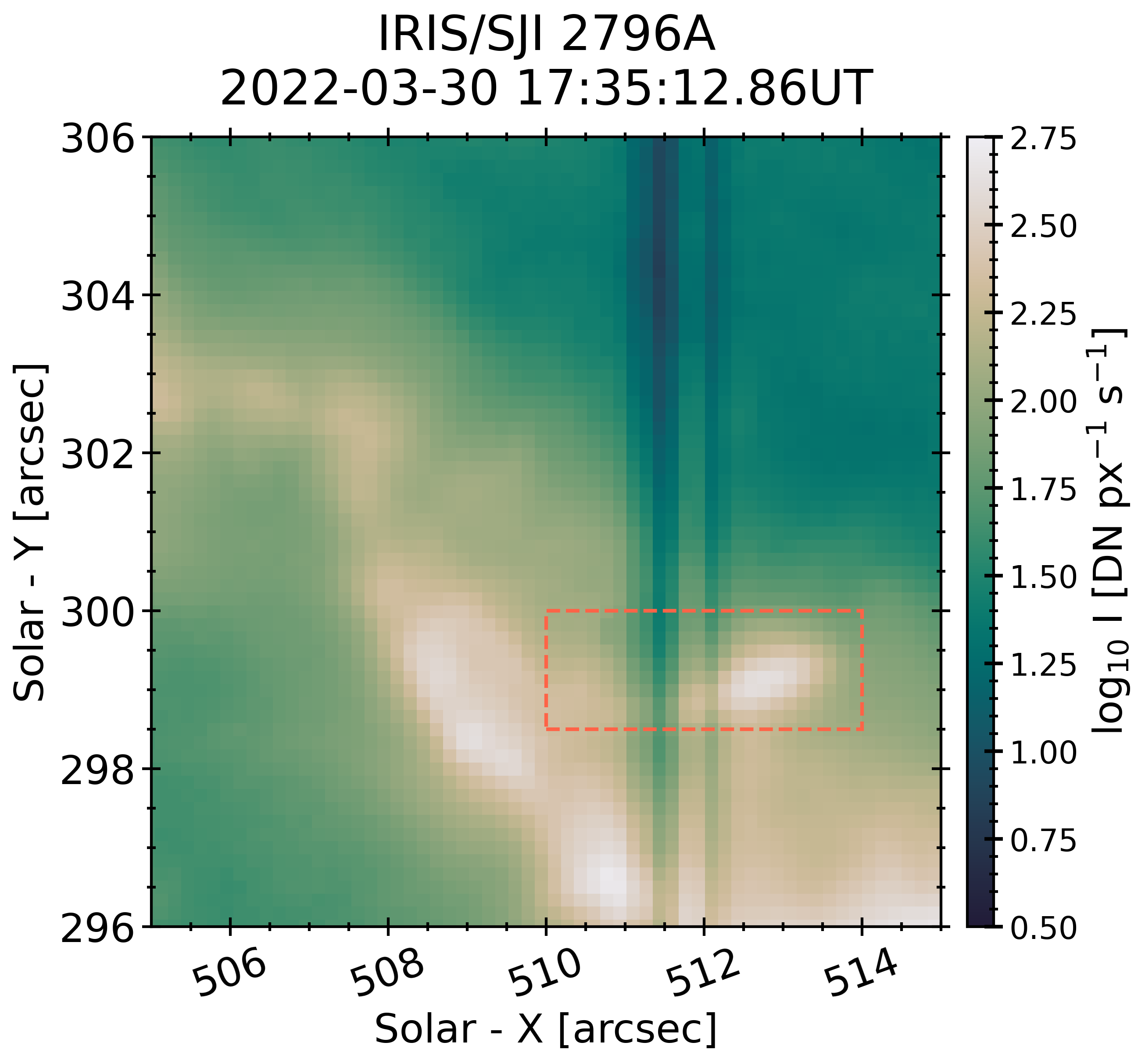}}
 \subfloat{\includegraphics[width = 0.315\textwidth, clip = true, trim = 0.cm 0.cm 0.cm 0.cm]{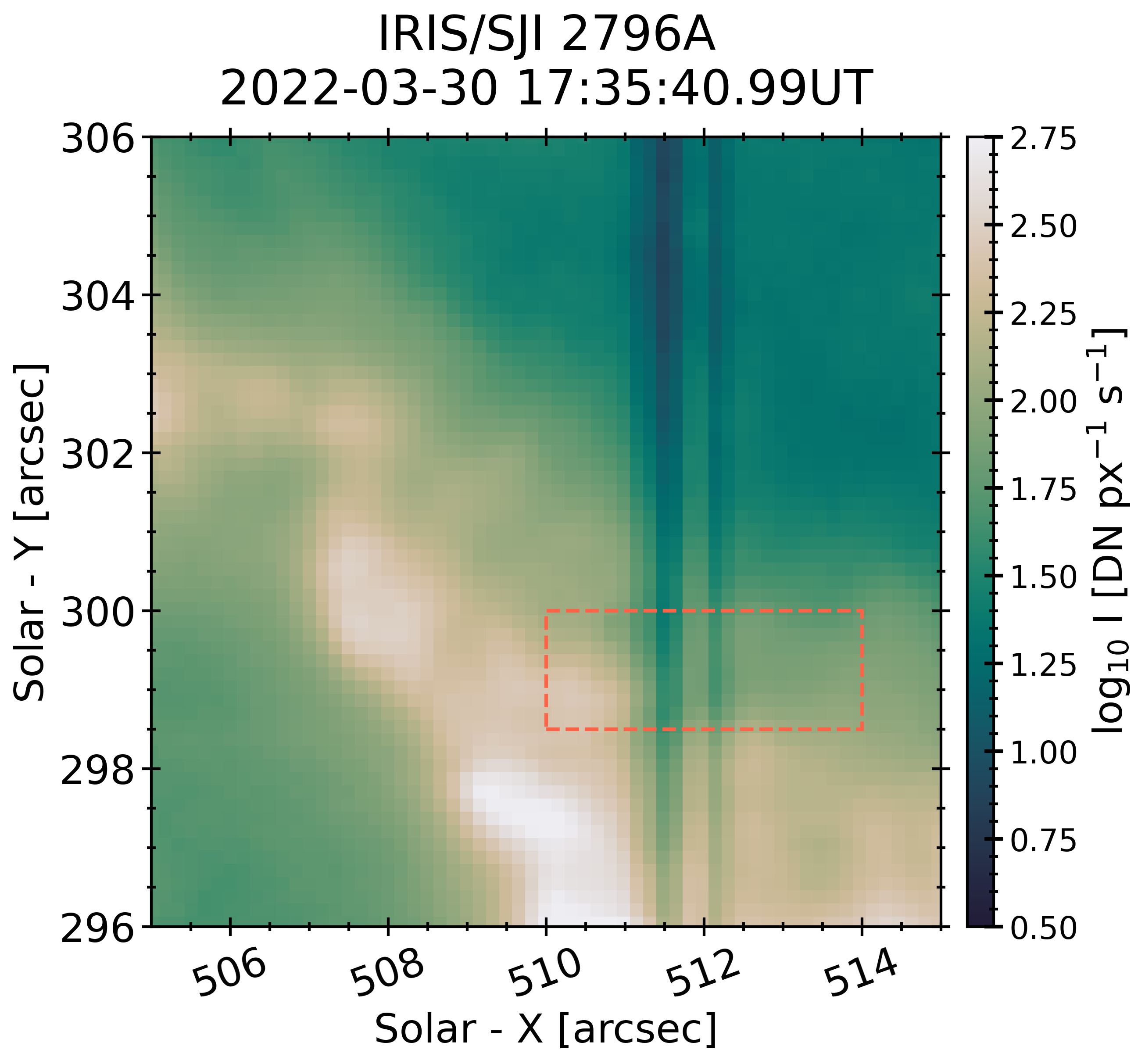}}
        }
        }
    \caption{\textsl{IRIS SJI images from the 2022-March-30th X-class flare, showing a sequence of $2796$~\AA\ observations. They show the initial brightening in the region of our source of interest, and then the re-brightening of this source a few minutes later. A rectangular box outlines the source of interest. The second brightening is associated with the appearance of extremely broad red-wings of several spectral lines. For one frame (at 17:35:19~UT) a small bright feature appears near x = 512\arcsec, y = 299\arcsec. By the next exposure it is no longer present. This feature could be a small extension of the southern ribbon, re-appearing in a region that was previously brightened. The darker vertical line in each image is IRIS' SG slit. The  distorted feature in the bottom of the image, extending from x = 510\arcsec-514\arcsec\ is the result of dust on the detector, which appears bright here since the image is shown on a logarithmic scale. Appendix~\ref{sec:sourceevol} shows a similar figure for the 1330~\AA\ filter.}}
	\label{Fig:SJIoverview_2796}
\end{figure*}

\section{Observations}
An X1.3 flare occurred around 17:21 UT on 2022 March 30, in active region (AR) 12975. It was well observed by several telescopes, such as IRIS, the Solar Dynamics Observatory's Atmospheric Imaging Assembly (SDO/AIA) and the Geostationary Operational Environmental Satellite (GOES). It was a typical two ribbon flare, with roughly north-south ribbon separation. IRIS observed this flare in sit-and-stare mode with a medium slit length (67\arcsec). Images were taken by IRIS's Slit-Jaw Imager (SJI) in the 1330~\AA, 1400~\AA\ and 2796~\AA\ passbands, with a cadence $\Delta t_\mathrm{SJI}\sim28$~s. Spectra were obtained with a variable cadence $\Delta t_\mathrm{SJI}\sim4-15$~s. The spatial resolution along the slit is $\sim$0.\arcsec33, and the spectral resolution in the near-UV is $\sim52$~m\AA\ and in the FUV is $\sim26$~m\AA. No on-board spectral summing was performed.

\begin{figure*}
	\centering
	\vbox{
	\hbox{
	\hspace{0.75in}
	\subfloat{\includegraphics[width = 0.75\textwidth, clip = true, trim = 0.cm 0.cm 0.cm 0.cm]{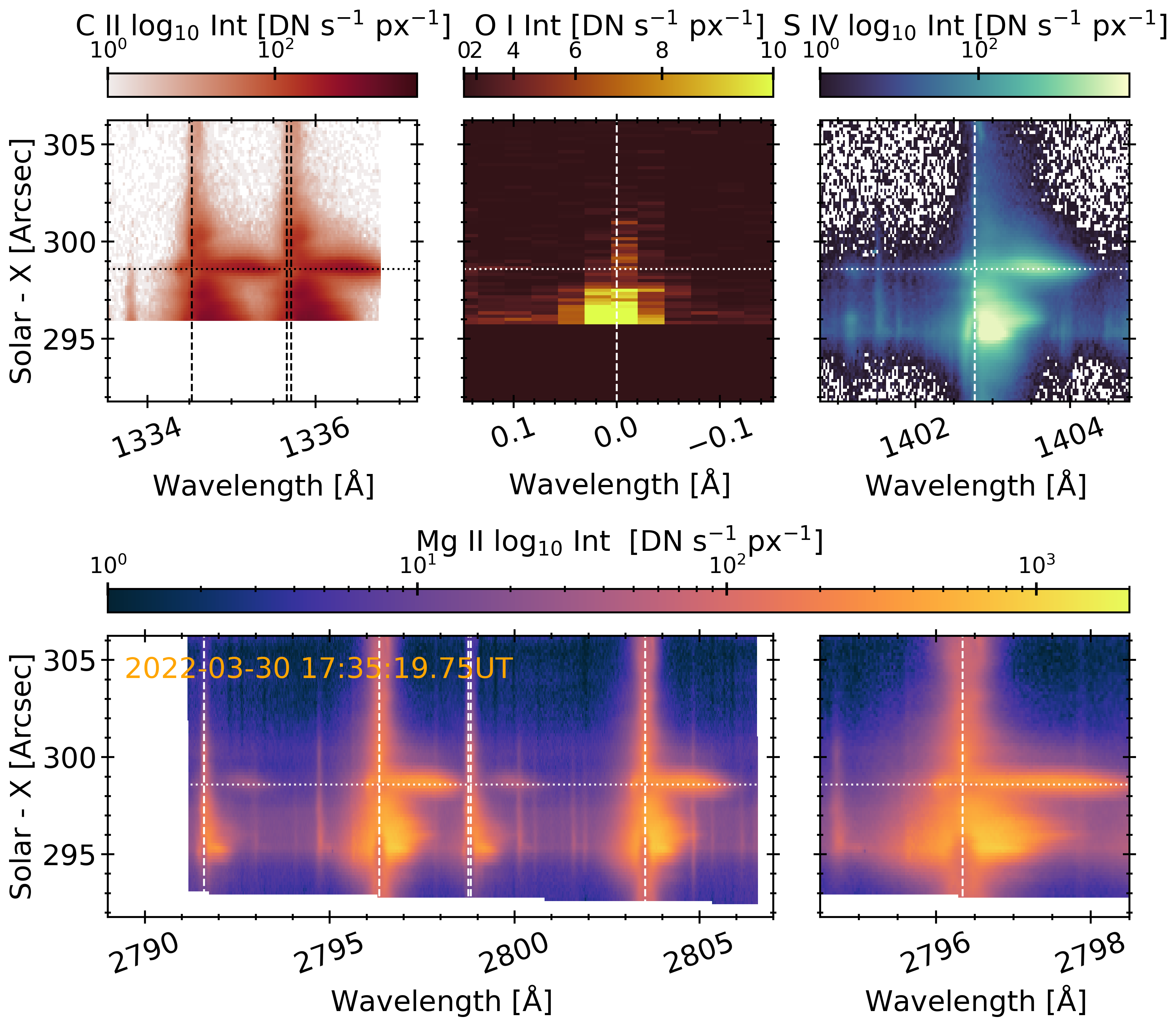}}
	}
	}
    \caption{\textsl{Spectra from the SOI, where the extremely broadened, redshifted wings are obvious. Shown are \ion{C}{2} (top left), \ion{O}{1} 1355.6~\AA\ (top-middle), \ion{Si}{4} (top right), \ion{Mg}{2} (bottom row, including a zoomed in view of \ion{Mg}{2} k). In each panel the vertical lines are the rest wavelengths of the main lines within each window. The dotted horizontal line indicates the location of the source of interest. Note that the \ion{O}{1} 1355.6~\AA\ does not exhibit a strong Doppler response within the SOI, but is very bright and broadened underneath the main ribbon, which is now in the southmost portion of the field of view. This line does increase a little in intensity after the SOI, perhaps in response to it.}}
	\label{Fig:SGoverview_SOI}
\end{figure*}

Figure~\ref{Fig:overview} provides an overview of this flare, showing snapshots of the flare ribbons in two SDO/AIA passbands and in a IRIS SJI passband, alongside the GOES soft X-ray lightcurves. A compact feature brightens behind the propagating bright flare ribbon, from which the unusually broad spectra are observed (see also Figure~\ref{Fig:SJIoverview_2796} \& \ref{Fig:SGoverview_SOI}). The source of interest (SOI) is approximately 5 IRIS pixels in size, which is about 1\arcsec.6. The emission from SOI is plotted in green in panel Figure~\ref{Fig:overview}(d). SOI has two peaks, the first peak is in response to the major flare ribbon and the second peak is caused by the transient heating being investigated in this study.

As reconnection proceeds along the arcade, there is an apparent motion of the two flare ribbons, with one sweeping south. This crosses the region near x = 510\arcsec, y = 298.5\arcsec first around 17:31 UT, shown in the top row of Figure~\ref{Fig:SJIoverview_2796}. Some 4-5 minutes later a small feature appears near x = 511\arcsec, y = 299\arcsec, in the wake of the ribbon, which itself continues to sweep southward. This small feature is the SOI, and appears to be an extension of the ribbon curving back in on itself. It persists for only SJI exposure. Appendix~\ref{sec:sourceevol} shows a similar figure for the 1330~\AA\ SJI filter.

\section{Extremely asymmetric red wings}

Spectrograms from the SOI, when the unusually broad spectra are observed, are shown in Figure~\ref{Fig:SGoverview_SOI} (Appendix~\ref{sec:prespectra} shows the spectra time at which the southern ribbon initially swept past the field of view). In each, the \ion{Si}{4}, \ion{C}{2}, and \ion{Mg}{2} resonance lines are shown, as is the \ion{O}{1} 1355.6~\AA\ line. When the ribbon initially passes over this region, around 17:31~UT, the spectral response is fairly typical of flare spectra, with broadened lines, redshifts and asymmetries, intensity increases, and single-peaked \ion{Mg}{2} resonance and subordinate line emission \citep[see, for example,][]{Kerr2015,Liu2015,Panos2018,2021ApJ...912..121P,2021ApJ...915...77P}. Later, during the SOI the spectra are quite atypical, with an obvious extreme excursion into the red wing of each of the transition region and upper chromospheric lines, that is the \ion{Si}{4} 1402~\AA\ line, the \ion{C}{2} resonance lines, the \ion{Mg}{2} h \& k lines, and the \ion{Mg}{2} subordinate triplet. At the peak time of the SOI, the spectra's red wings are fairly flat, and can even exceed the intensity of the core region. Excursions into the red wing are so extreme that the \ion{C}{2} lines merge into one another, and the \ion{Mg}{2} k line intersects the blue wing of the \ion{Mg}{2} 2799~\AA\ subordinate lines. There are also times when the \ion{Mg}{2} resonance lines have a shallow central reversal. Notably \ion{O}{1} 1355.6~\AA\, the core of which forms somewhat deeper than the \ion{Mg}{2} lines \citep{Kerr2019a}, does not respond very strongly at the time of the SOI, nor does it exhibit the red wing asymmetry. Similarly, \ion{O}{4} 1401.163, which forms at similar temperatures to \ion{Si}{4}, remains fairly weak. The O lines are optically thin, which might suggest opacity effects are in some way responsible for the red wing characteristics of the other spectra. Overall, the intensity for all of the lines is smaller than the initial brightening. For the remainder of this analysis we focus on the evolution of a single pixel, which exhibited the broadest red wings, at index $y_{\mathrm{slit}} = 51$.

Due to the relatively coarse temporal resolution of SJIs, we estimate the duration of SOI using the \ion{Mg}{2} data. The red-wing enhancement becomes obvious at 17:34:48 UT and disappears by 17:35:29 UT. The time intervals immediately before and after this 41~s period  are 17:34:33 UT and 17:35:44 UT. Therefore, the estimated duration of this transient feature, to within the cadence of the observations, is $41 \pm 15$~s.

\begin{figure*}
	\centering
	\vbox{
	\hbox{
	\subfloat{\includegraphics[width = 0.425\textwidth, clip = true, trim = 0.cm 0.cm 0.cm 0.cm]{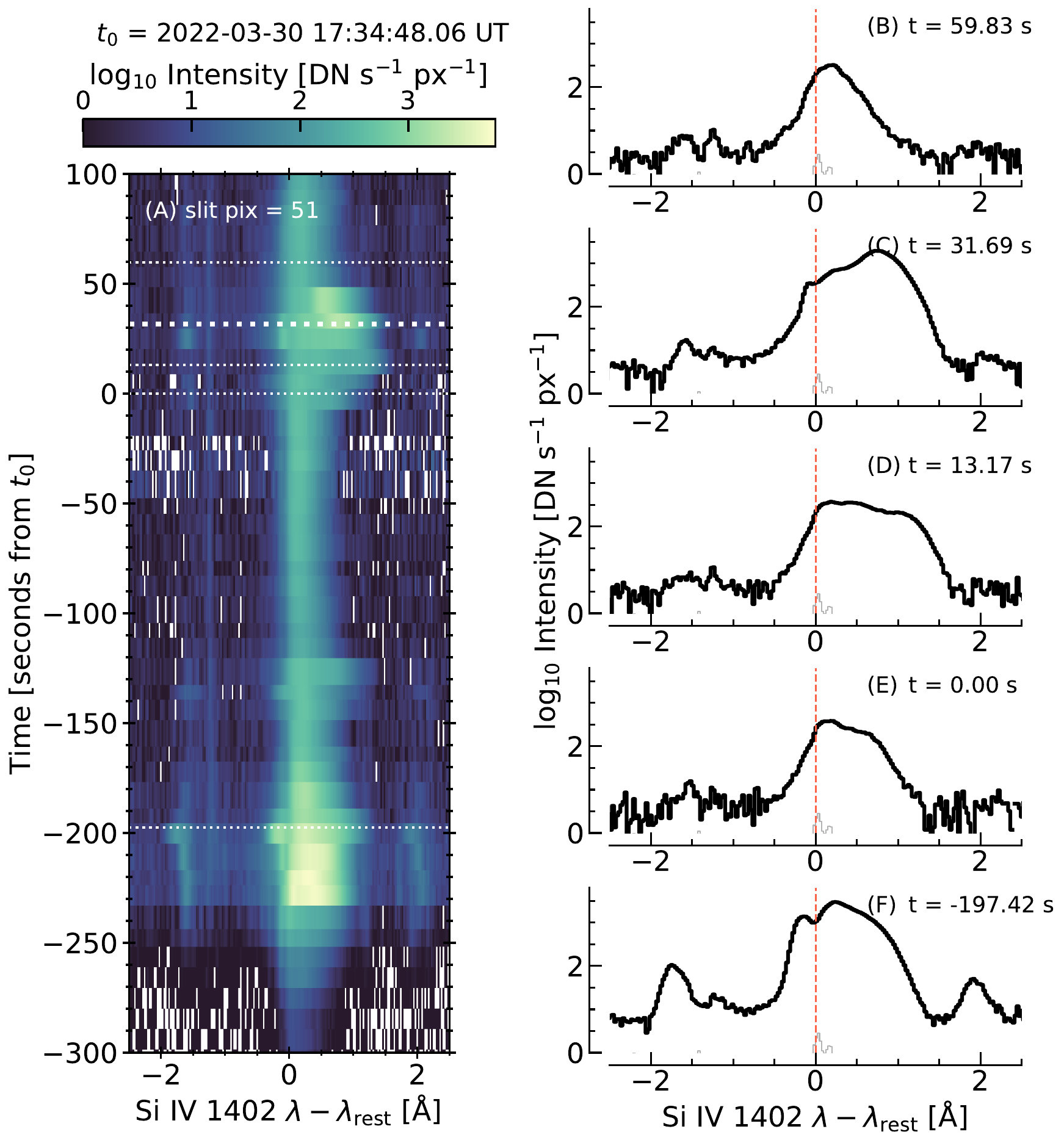}}
 	\subfloat{\includegraphics[width = 0.425\textwidth, clip = true, trim = 0.cm 0.cm 0.cm 0.cm]{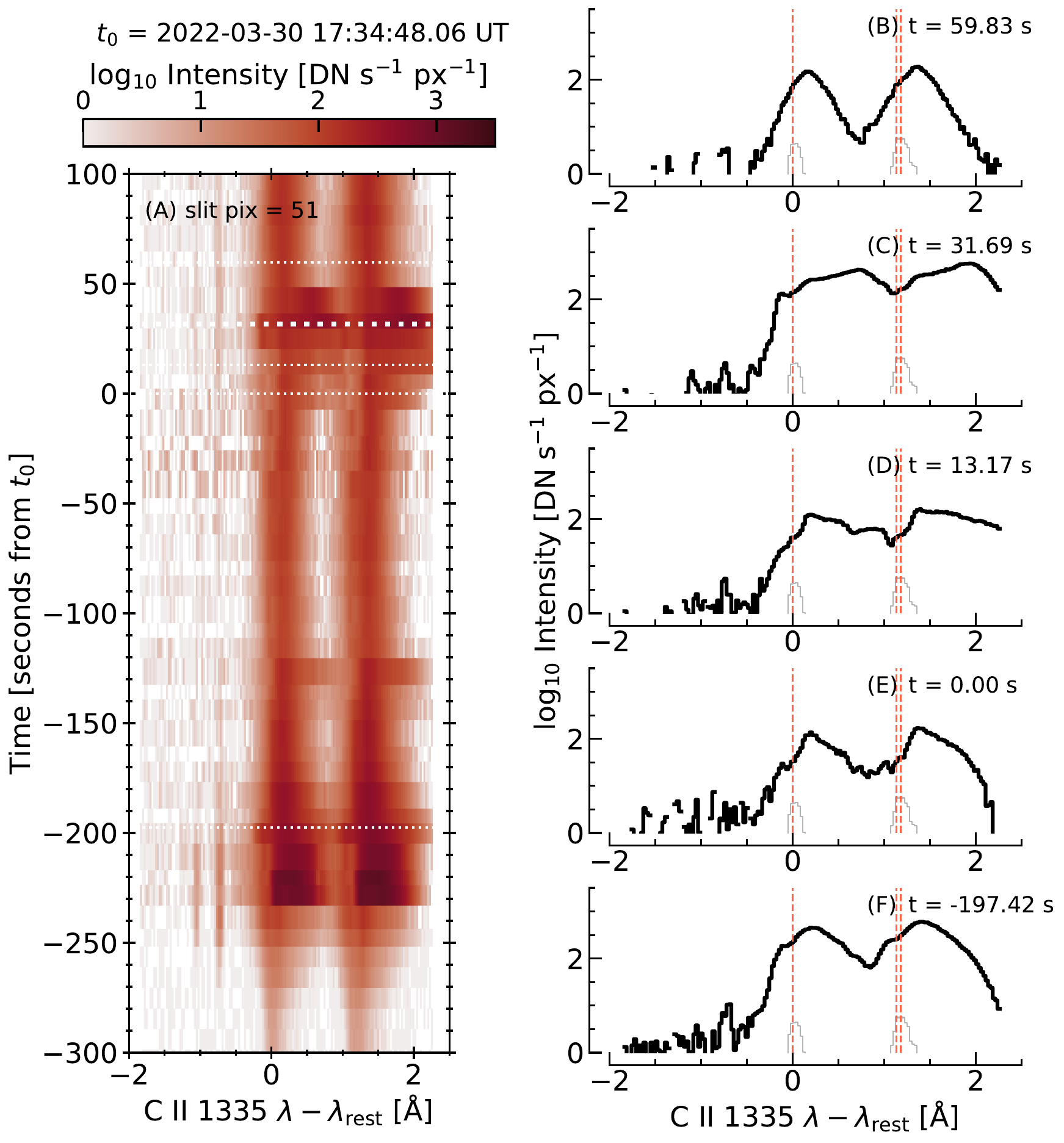}}
         }
         }
        \vbox{
	\hbox{
        \subfloat{\includegraphics[width = 0.425\textwidth, clip = true, trim = 0.cm 0.cm 0.cm 0.cm]{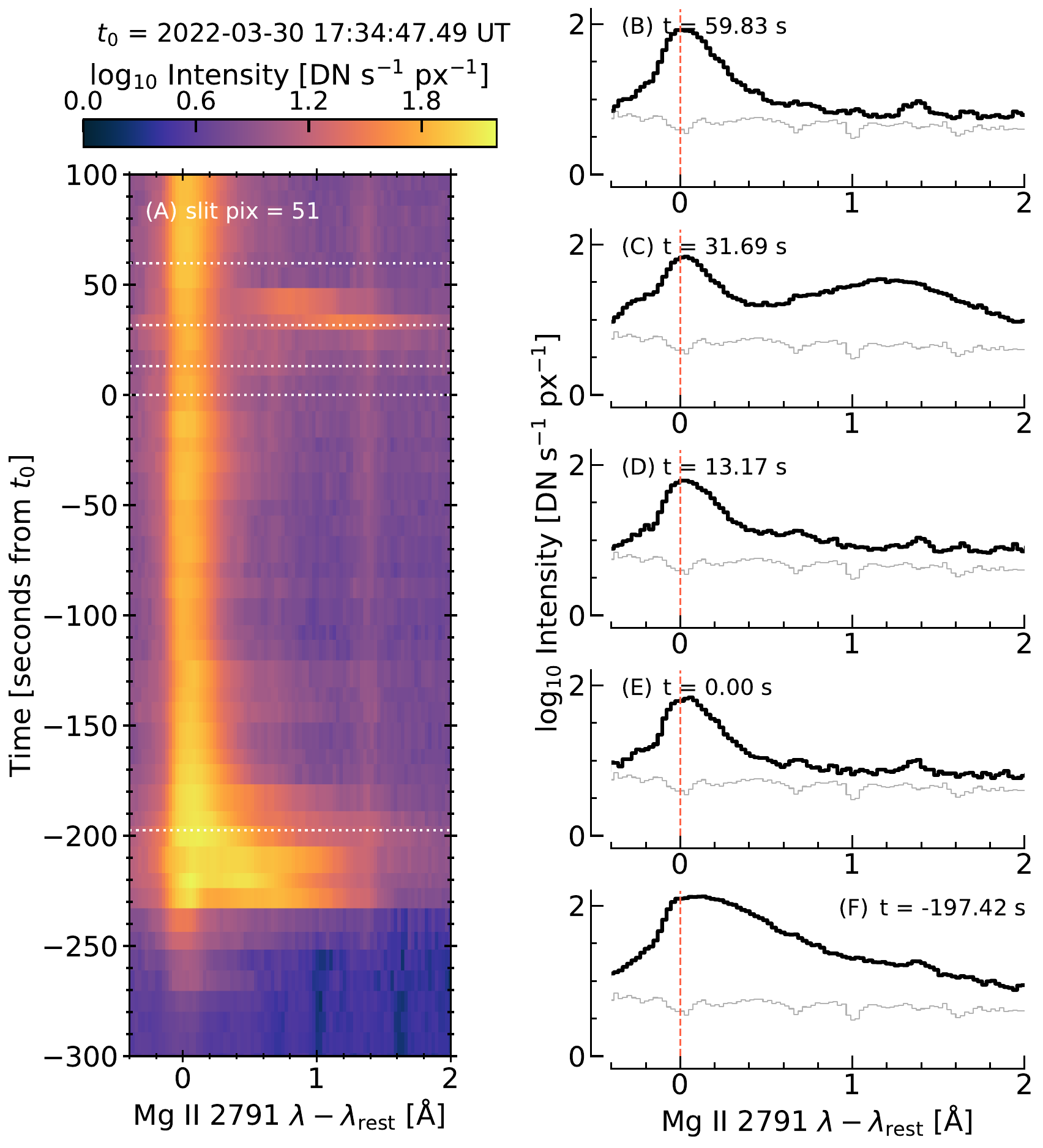}}
         \subfloat{\includegraphics[width = 0.425\textwidth, clip = true, trim = 0.cm 0.cm 0.cm 0.cm]{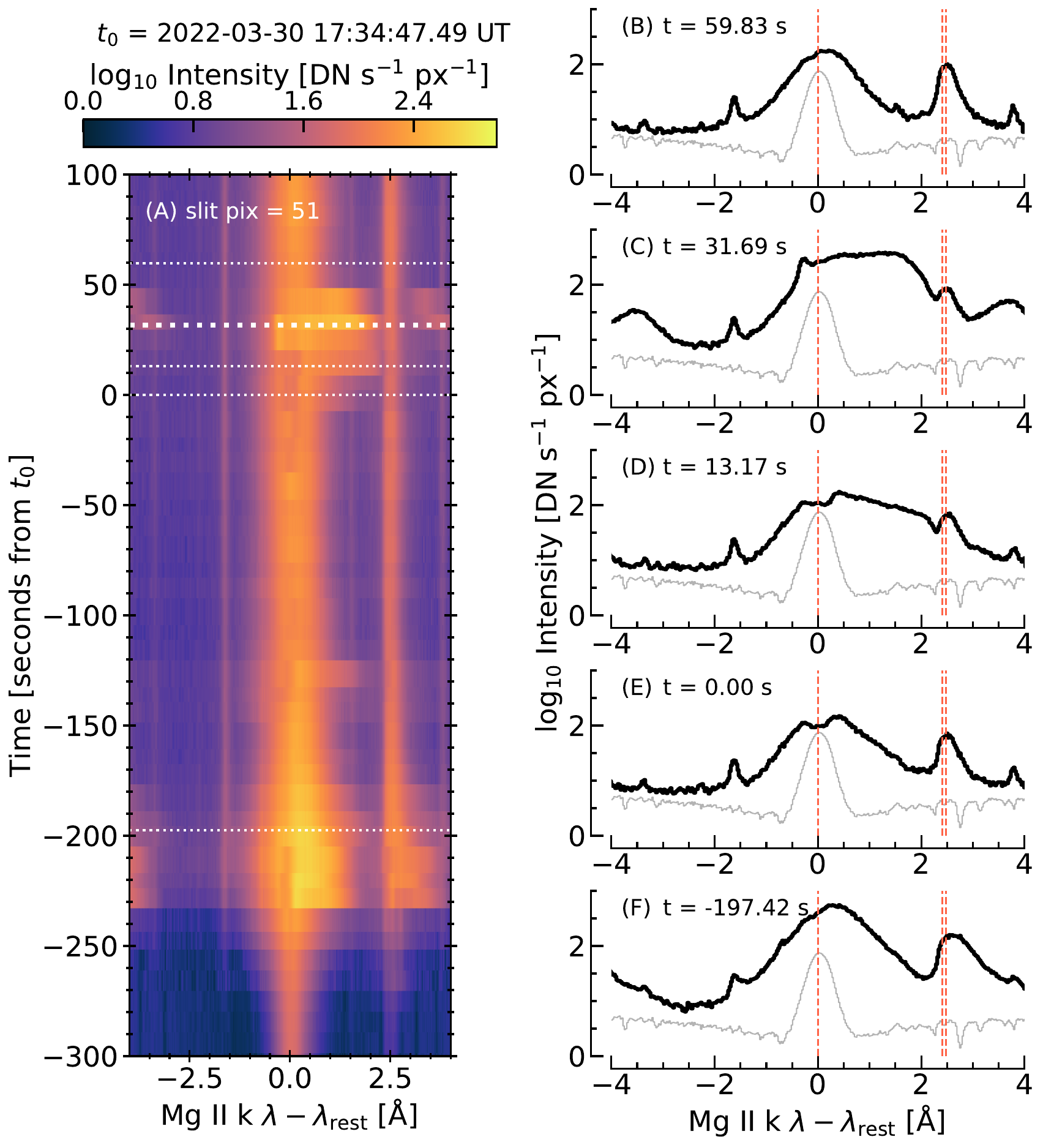}}
	}
	}
    \caption{\textsl{Temporal evolution of the \ion{Si}{4} (top left), \ion{C}{2} (top right; note that all three \ion{C}{2} resonance lines are included within this window, two of which are blended}), \ion{Mg}{2} 2791~\AA\ subordinate line (bottom left), and \ion{Mg}{2} k (bottom right). In each the image represents wavelength-versus-time, where the zero point is indicated in the figure titles. Cut-outs at various times are shown also (indicated by the horizontal lines; the thicker horizontal lines are the times for which example Gaussian fits are shown in later figures). In the cut-outs the red vertical lines are the rest wavelengths of the strong lines in each passband. The reference time ($t_{0}$) was selected to be the first instance when the red wing of the \ion{Mg}{2} resonance lines became broadened during the second brightening of the source of interest.}
	\label{Fig:wavetime}
\end{figure*}

The evolution of the atypical spectra as a function of time are shown in detail in Figure~\ref{Fig:wavetime}, where the image of each spectral line shows wavelength as a function of time, and where the cut-outs show the spectra at various times. In those figures the rest wavelengths are indicated (note that one of the \ion{C}{2} resonance lines is a blend of two strong lines, and similarly the \ion{Mg}{2} subordinate triplet includes a blend of two of those lines near 2799~\AA).

The three resonance line spectra exhibit very similar line shapes during the SOI: the red wings grow in intensity, flatten, and extend over 200~km~s$^{-1}$ from the rest wavelength, without an equivalent feature in the blue wing. As mentioned previously, these are very transient features, with a lifetime less than 60~s. Inferred flows are discussed more quantitatively later, but we note here that these red wings imply supersonic flows (the sound speed is likely $v_{s}< 50$~km~s$^{-1}$ at the formation temperature of the lines considered here) that appear within $15$~s and dissipate rapidly, within $<55$~s.

While the three resonance lines seemingly brighten and exhibit enhanced red wings simultaneously (within the $\Delta t = 4-15$~s cadence), the weaker, and intrinsically narrower, \ion{Mg}{2} subordinate triplet seem to take somewhat longer to exhibit the red wing asymmetry, though there is a slight increase in the continuum level which could be hiding a weak asymmetry. Some 30~s after the resonance lines clearly show an excursion into the red wing, one appears in the triplet lines also. This appears as a separate component, that is weaker than the stationary component. Differences between the appearance of this component compared to the same feature in the resonance lines could be due to the relative narrowness of the triplet lines, or due to their slightly deeper formation height. This separate red-shifted component analogous to, but more extreme than, the IRIS spectra studied by \cite{Graham2020}. In that study, which was backed up by radiation hydrodynamic modelling, the authors attributed the appearance of a separate component that decelerated towards rest, ultimately merging with the stationary component, to a dense chromospheric condensation. The temporal evolution of our SOI is not immediately clear from the wider resonance line spectra, but there are hints from the narrower subordinate lines spectra that we are seeing a similar pattern as reported by \cite{Graham2020} (c.f. lower left hand panel of Figure~\ref{Fig:wavetime}).

The initial brightening is also shown on Figure~\ref{Fig:wavetime}, around $t \sim -225$~s, for comparison. The peak intensity of some of the lines is larger during the initial brightening, but while the wings do broaden, they do not take on the flat shape and do not extend as far as they do during the SOI. Also the \ion{Mg}{2} triplet line wings appear as smooth extensions rather than a separate component.

\begin{figure}
	\centering
	\vbox{
	\hbox{
	\subfloat{\includegraphics[width = 0.425\textwidth, clip = true, trim = 0.cm 0.cm 0.cm 0.cm]{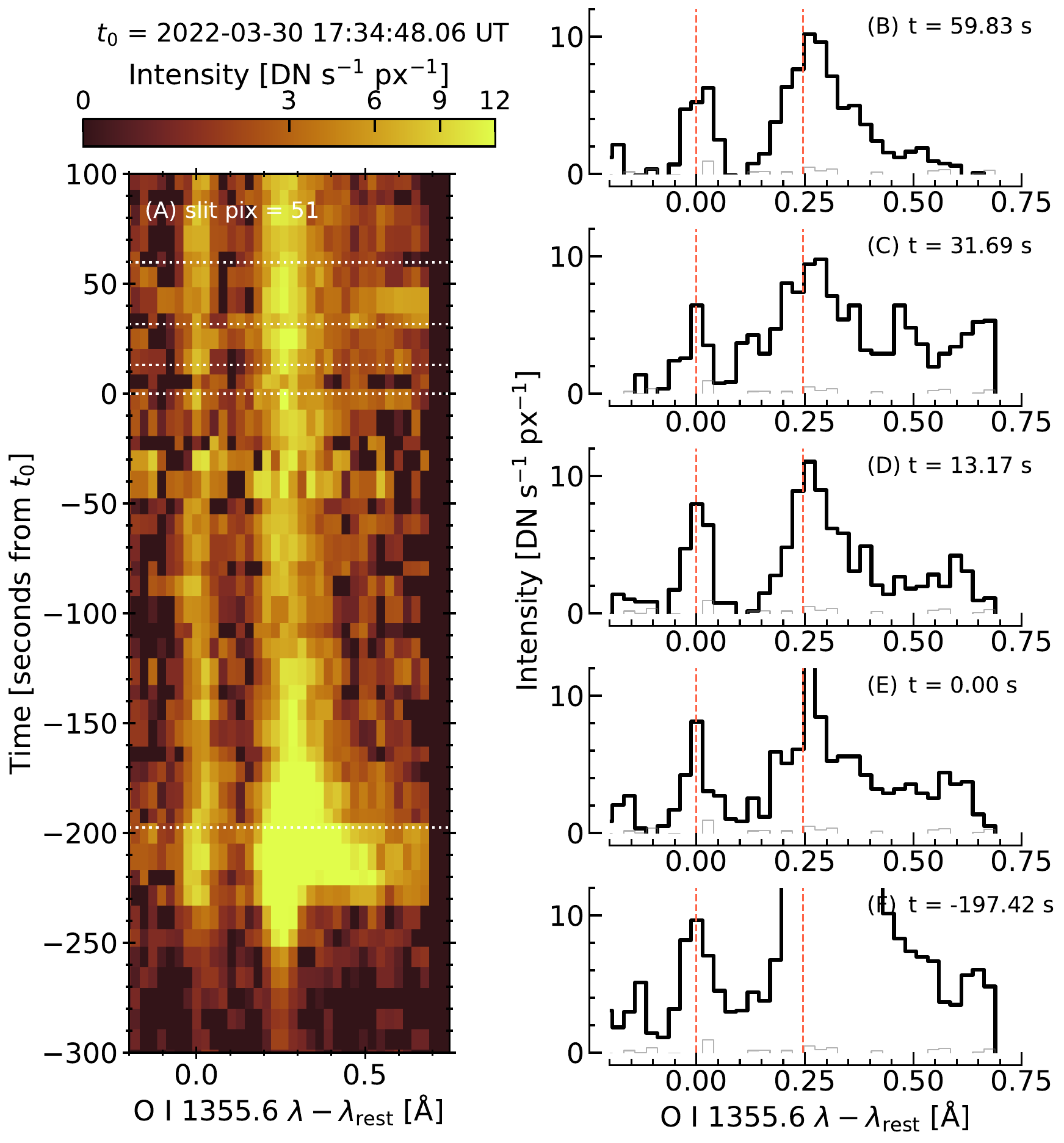}}
         }
         }
         \vbox{
	\hbox{
	\subfloat{\includegraphics[width = 0.425\textwidth, clip = true, trim = 0.cm 0.cm 0.cm 0.cm]{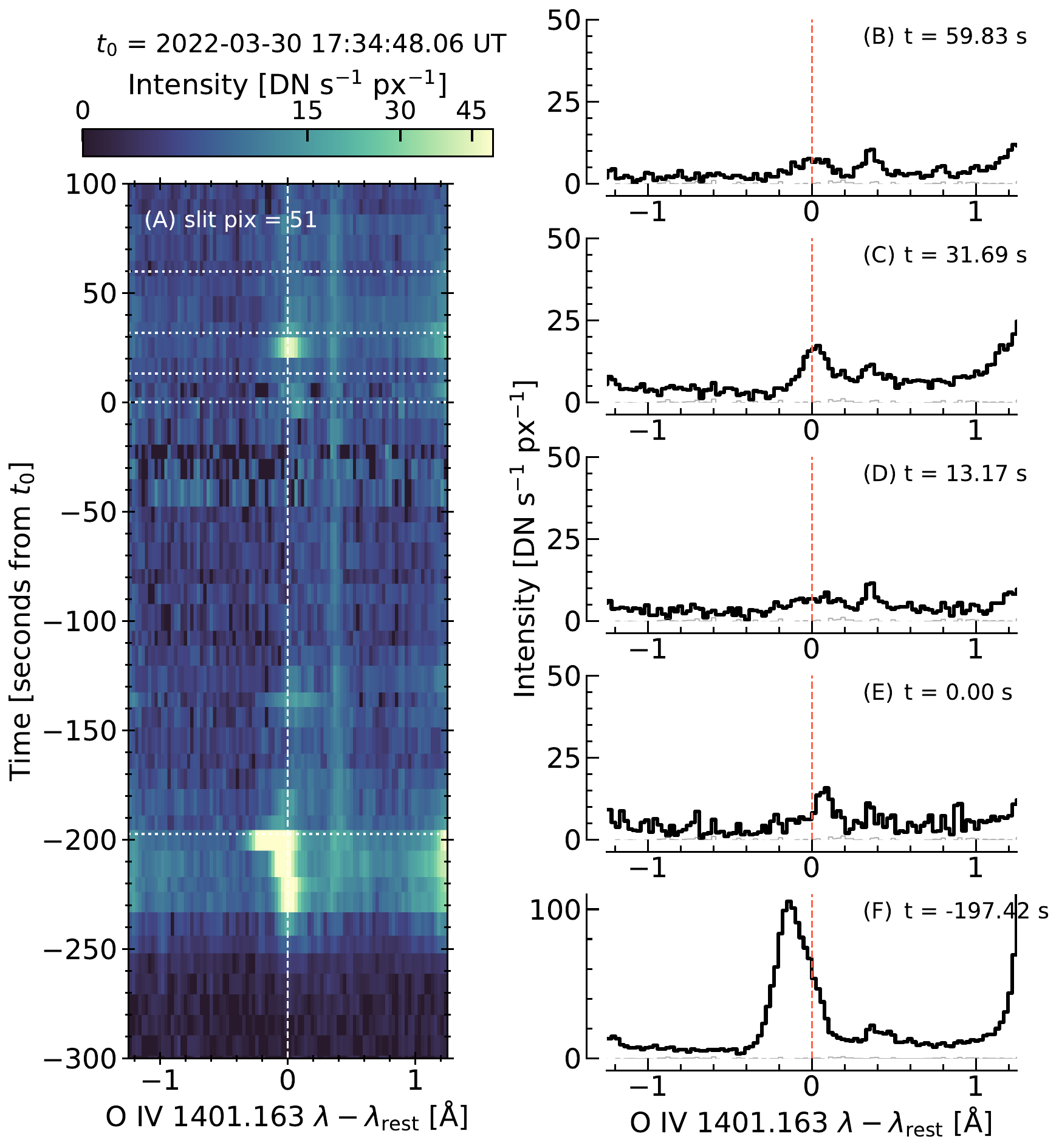}}
         }
         }
    \caption{\textsl{Same as Figure~\ref{Fig:wavetime}, but for the \ion{O}{1} 1355.598~\AA\ and \ion{C}{1} 1355.884~\AA\ lines (top panel), and \ion{O}{4} 1401.163~\AA\ (bottom panel).}}
	\label{Fig:wavetime_oi}
\end{figure}

Figure~\ref{Fig:wavetime_oi} shows the temporal evolution of the \ion{O}{1} 1355.598~\AA\ line, with the \ion{C}{1} 1355.844~\AA\ line appearing adjacent to it. Unlike the lines that form at somewhat greater altitude and at somewhat higher temperature, this line exhibits no extraordinary red wing asymmetries, and remains relatively weak. There are hints of red wing asymmetries in the \ion{C}{1} line, but they are not as noteworthy as the other spectra. A lack of a red-wing feature in the \ion{O}{1} 1355.598~\AA\ line strongly implies that this unusual behaviour is confined to the transition region and upper chromosphere. Also shown on Figure~\ref{Fig:wavetime_oi}  is the \ion{O}{4} 1401.163~\AA\ line, which shows a modest intensity increase during the SOI, but does not exhibit any significant broadening or Doppler flows. This line peaks in intensity one exposure \textsl{prior} to the appearance of the very flattened wings of the resonance lines. During the initial brightening the \ion{O}{4} line is blueshifted, suggestive of upflowing plasma. Under the assumption of ionisation equilibrium \ion{Si}{4} forms at $T\sim80$~kK, and \ion{O}{4} forms at $T\sim140$~kK. The lack of major response in during the SOI restricts any substantial downward mass motions to $T< \sim140$~kK.  Several effects, including non-equilibrium ionization \citep{Doyle13}, non-Maxwellian distributions \citep{Dudik14} and suppression of dielectronic recombination \citep{Polito16b} can shift the formation temperature of these lines, to either higher or lower values.

\section{Results of Doppler Velocities}

There are several methods that can be used to obtain quantitative information from spectra, which can be used to infer some plasma properties or to compare to numerical models. To find the Doppler shift, a common method is to fit the spectra with a single or multiple Gaussian functions \citep[recent example with IRIS data include][]{Li2022, Yu2020, Ashfield2022}. This method works well for optically thin lines or spectra with Gaussian shapes. On the other hand, spectral moments analysis is appropriate for optically thick lines, such as the \ion{Mg}{2} lines, or line profiles that differ from Gaussian shape. The line centroid is defined as $\lambda_c = M_{1}/M_{0}$, where
$M_{0} = \int I(\lambda)\mathrm{d}\lambda$ (0th moment) and $M_{1} = \int I(\lambda)\lambda\mathrm{d}\lambda$ (1st moment). The Doppler shift can then be found by measuring the displacement between this centroid and the reference line center (at a non-flare region). In principle, the method using spectral moments provides a sense of an ``averaged'' Doppler signal, since it measures the `center of mass' of the overall line.

\begin{figure}[bh]
\vbox{
\hbox{
\subfloat{\includegraphics[width = 0.325\textwidth, clip = true, trim = 0.cm 0.cm 0.cm 0.cm]{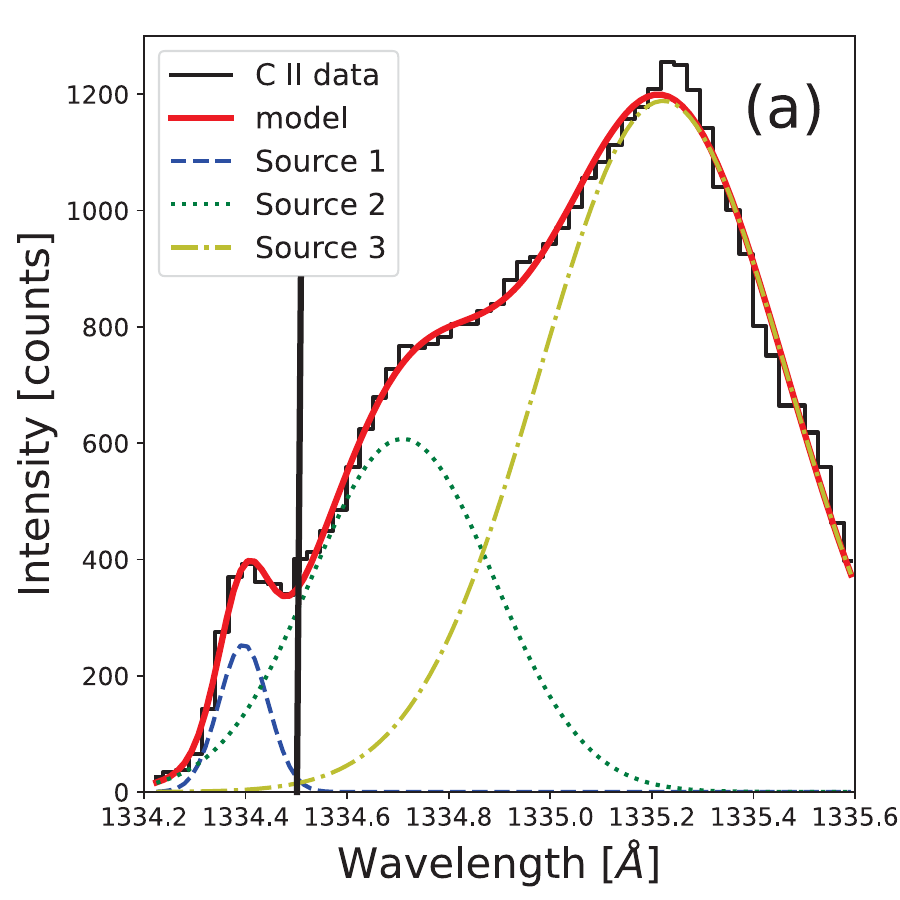}}
}
}
\vbox{
\hbox{
\subfloat{\includegraphics[width = 0.325\textwidth, clip = true, trim = 0.cm 0.cm 0.cm 0.cm]{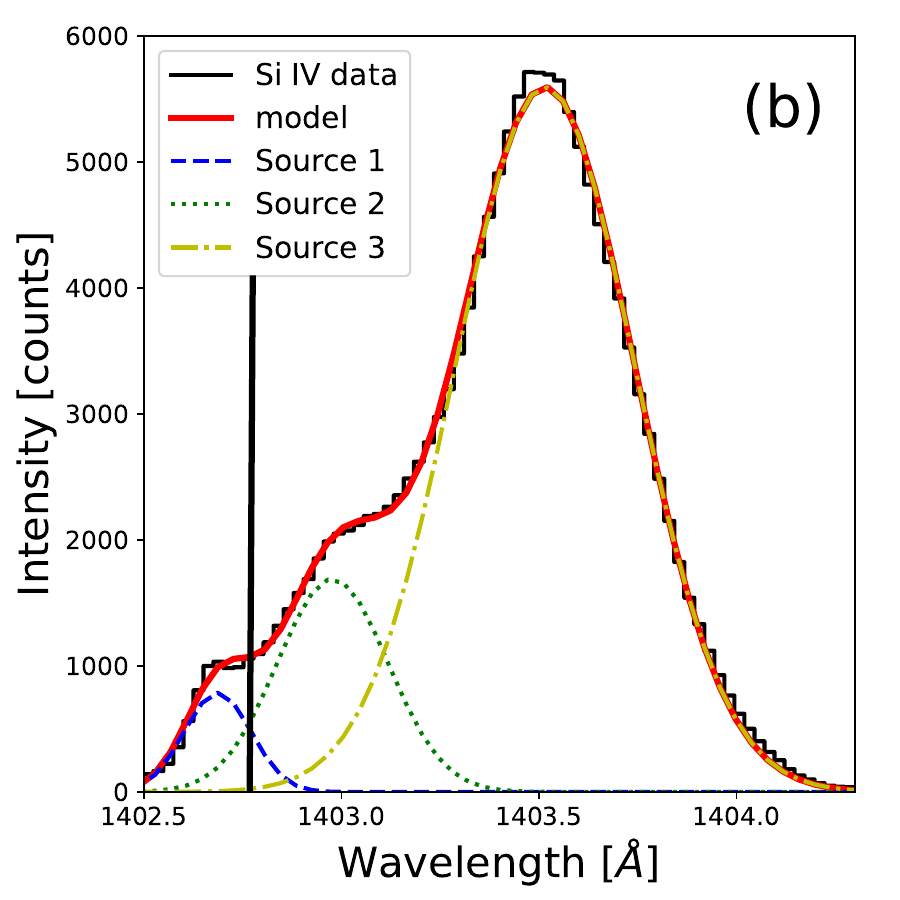}}
}
}
\vbox{
\hbox{
\subfloat{\includegraphics[width = 0.325\textwidth, clip = true, trim = 0.cm 0.cm 0.cm 0.cm]{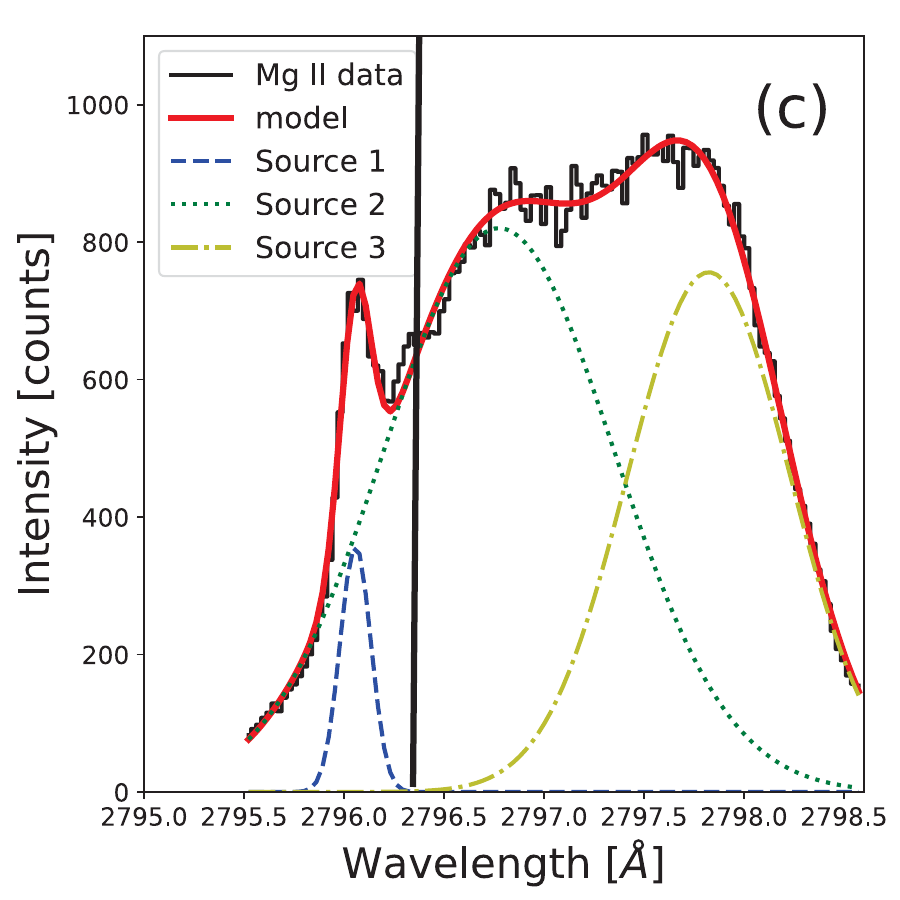}}
}
}
\caption{\textsl{Multi-Gaussian fit of UV spectra at 17:35:19 UT on SOI (Y = 51 pixel). Three Gaussian components are involved and the combinations of all components are the modeled results in red solid curves. The dotted-dash line in yellow is the red-most Gaussian fit, representing the downward Doppler velocities. The relative timing of these spectra are indicated by thick horizontal lines on Figure~\ref{Fig:wavetime}. The vertical line in each panel shows the rest line center positions of \ion{C}{2}, \ion{Si}{4} and \ion{Mg}{2}, at 1334.53~\AA, 1402.77~\AA, and 2796.33~\AA, respectively.}}
\label{Fig:measure}
\end{figure}

\begin{figure}[h]

\vbox{
\hbox{
\subfloat{\includegraphics[width = 0.425\textwidth, clip = true, trim = 0.cm 0.cm 0.cm 0.cm]{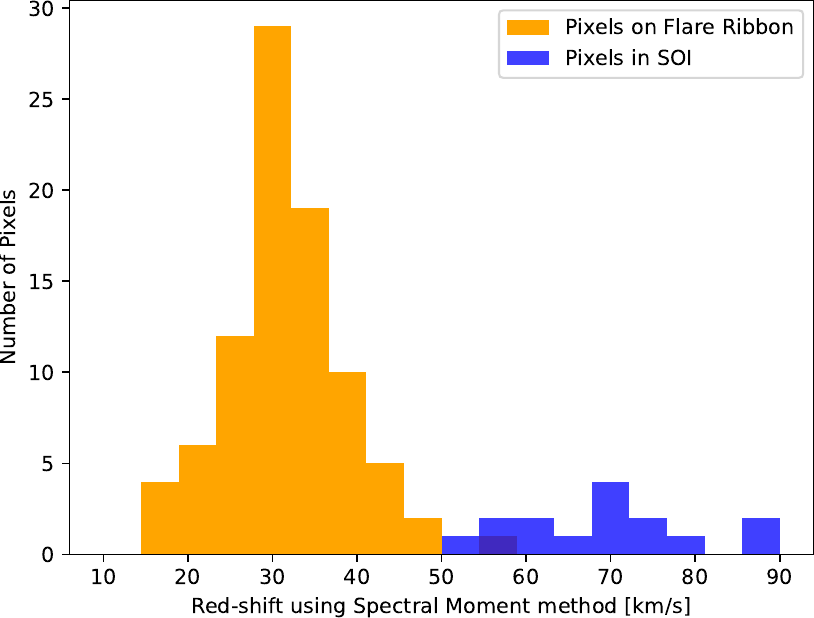}}
}
}
\caption{\textsl{Histogram of \ion{Mg}{2} red-shifts of pixels on flare ribbon (orange) and SOI region (blue).}}
\label{Fig:hist}
\end{figure}

\begin{figure}[bh]
\vbox{
\hbox{
\subfloat{\includegraphics[width = 0.45\textwidth, clip = true, trim = 0.cm 0.cm 0.cm 0.cm]{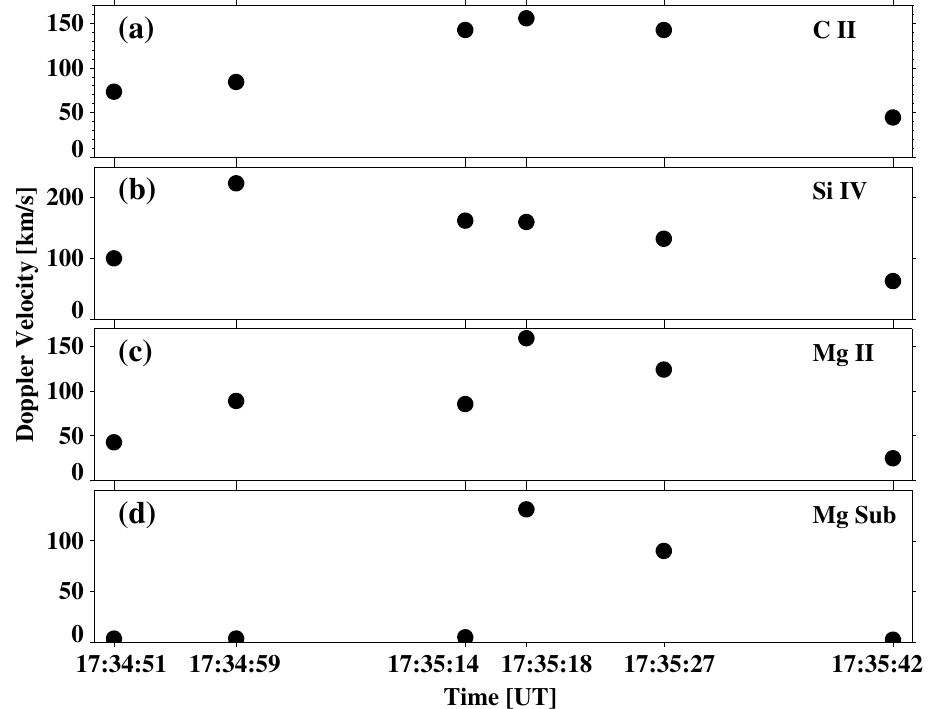}}
}
}
\vbox{
\hbox{
\subfloat{\includegraphics[width = 0.45\textwidth, clip = true, trim = 0.cm 0.cm 0.cm 0.cm]{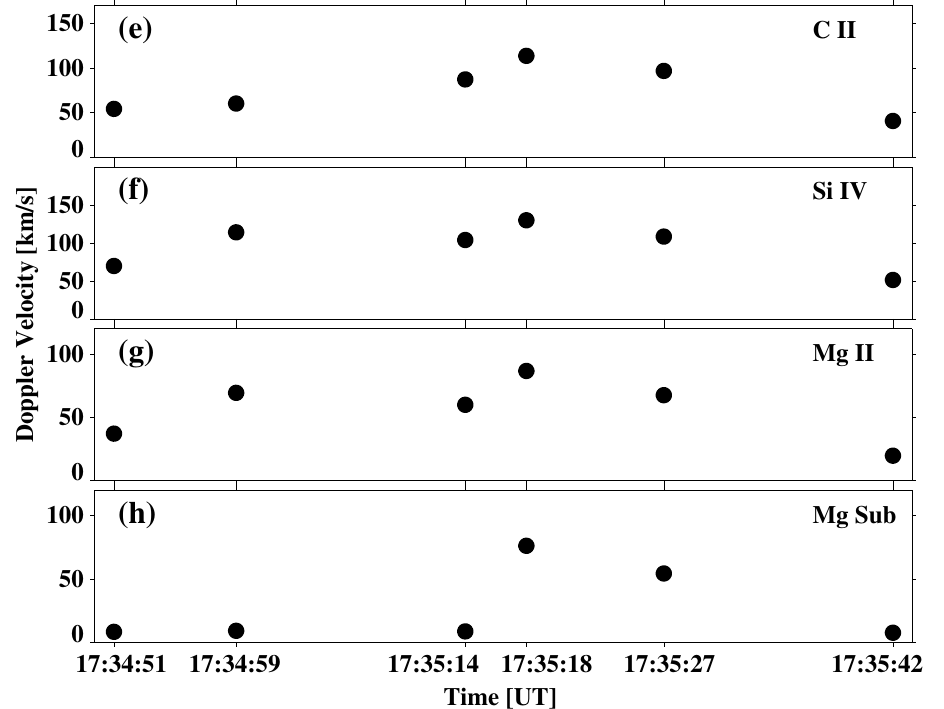}}
}
}
\caption{\textsl{Evolution of derived red-shift velocities, using multi-Gaussian fit in panels (a) - (d) and spectral moment method in panel (e) - (h),  on SOI (Y = 51 pixel). }}
\label{Fig:timeseq}
\end{figure}

Figure~\ref{Fig:measure} shows UV spectra at the time of peak asymmetry in the SOI  (17:35:19.75 UT), for the \ion{C}{2} 1334.53~\AA\, \ion{Si}{4} 1402.77~\AA\ line, and the \ion{Mg}{2} k line. Three Gaussian components are used in the fitting and the representative downward velocity was obtained from the red-most Gaussian fit (dotted dash curves). The results are shown in Table~\ref{results}, alongside the moments analysis. The multi-Gaussian fits suggest a large-scale motion with a speed about 160 km s$^{-1}$ in the upper chromosphere and transition region. During the peak time (17:35:19 to 17:35:29 UT), the red-wing component of the \ion{Mg}{2} subordinate lines is present. These lines form somewhat below the \ion{Mg}{2} h \& k lines during flares \citep[e.g.][]{Zhu2019,Kerr2019a,Kerr2019b}, much higher in altitude than in the non-flaring atmosphere. Their Doppler shift at  17:35:19 was measured to be 130  km s$^{-1}$. The moments analysis yields smaller Doppler shifts for each line than the results obtained by Gaussian fitting, but are still larger than typically observed flare sources. A histogram of \ion{Mg}{2} Doppler shifts derived from the moments analysis including pixels in both SOI (blue) and the main flare ribbon (orange) is shown in Figure~\ref{Fig:hist}. On the flare ribbon (pixels 27 to 48), the peak shift is around 30~km~s$^{-1}$. Most of the redshifts of SOI pixels range from 50 to nearly 90~km~s$^{-1}$. For those pixels in the SOI with large moments-derived Doppler shifts our Gaussian fitting analysis (c.f Figures~\ref{Fig:measure} \& \ref{Fig:timeseq}) yields Doppler motions far into the wings, up to 160~km~s$^{-1}$. The temporal evolution of the Doppler shifts from the SOI in different spectral lines are plotted in Figure~\ref{Fig:timeseq}. Most lines reached their peak redshift around 17:35:19 UT, except the red-most Gaussian fit of \ion{Si}{4} line shows a peak velocity of about 220 km s$^{-1}$ at 17:34:59 UT using the multi-Gaussian fit.

While interpreting velocity evolution in optically thick lines is non-trivial, we note the following regarding Figure~\ref{Fig:timeseq}. There is general agreement in both order of magnitude of the inferred Doppler shifts from each of our methods, as well as with the temporal evolution after 17:35:14~UT. In the 20-30~s following the peak, the inferred Doppler shifts decay towards rest, at which times the excessively broadened red wings becoming less prominent. The lines later appear broadened and slightly asymmetric but more like typical flare profiles. This timescale is consistent with the time taken for the separable red wing components observed by \cite{Graham2020} in flare spectra to merge with the stationary components. Indeed, this is most easily seen in the narrower \ion{Mg}{2} 2791~\AA\ line, where there does appear to be a separable component that merges with the stationary component (c.f. Figure~\ref{Fig:wavetime}, lower left panels). The stronger opacity and larger widths in the other lines may act to mask the shifted component as being a distinct separate component from the stationary profiles. While the initial shifts were not as large as those inferred in this flare, \cite{Graham2020} attributed such behaviours to the development, propagation, and deceleration of chromospheric condensations.

\begin{table}[h]
\begin{center}
\caption{Downward Doppler Velocity Measured by Different Methods, at the peak time of 17:35:19 UT}
\label{results}

\begin{tabular}{lcc}
\\
 \hline\noalign{\smallskip}
Spectral Lines  &	    Spectral Moment	    &   Multi-Gaussian \\
 \hline\noalign{\smallskip}
\ion{C}{2}	    &		115 km~s$^{-1}$	    &	155 km~s$^{-1}$	 \\
\ion{Si}{4}		&		125 km~s$^{-1}$	    &	160 km~s$^{-1}$	 \\
\ion{Mg}{2}		&	    85 km~s$^{-1}$	    &	160 km~s$^{-1}$	 \\
\ion{Mg}{2} subordinate &  75 km~s$^{-1}$	&   130 km~s$^{-1}$	 \\

 \hline\noalign{\smallskip}\hline
\end{tabular}
\end{center}
\end{table}

\section{Discussion}

\begin{figure}[bh]

\vbox{
\hbox{
\subfloat{\includegraphics[width = 0.425\textwidth, clip = true, trim = 0.cm 0.cm 0.cm 0.cm]{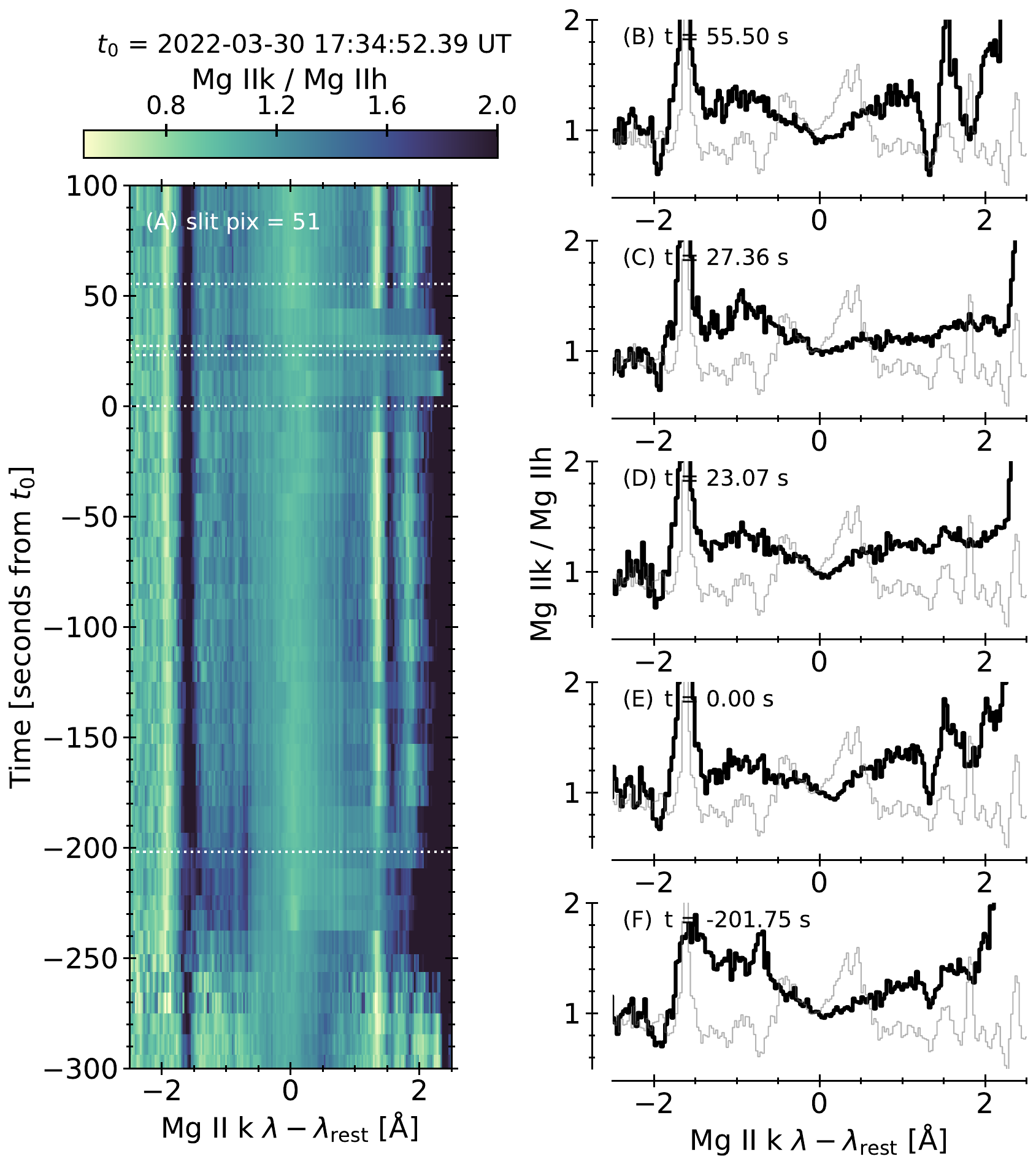}}
}
}
\caption{\textsl{The evolution of the \ion{Mg}{2} k:h ratio within the source of interest as a function of time. The image shows wavelength-versus-time, where the decrease in the ratio closer to unity in the red wing occurs near $t=0$~s. The blue wing has a somewhat higher ratio. This can also be seen in the cutouts at specific times. The initial flare brightening produces a similar effect, but does not penetrate as far into the red wing.}}
\label{Fig:wavetime_asymm}
\end{figure}

\begin{figure}[h]

\vbox{
\hbox{
\subfloat{\includegraphics[width = 0.425\textwidth, clip = true, trim = 0.cm 0.cm 0.cm 0.cm]{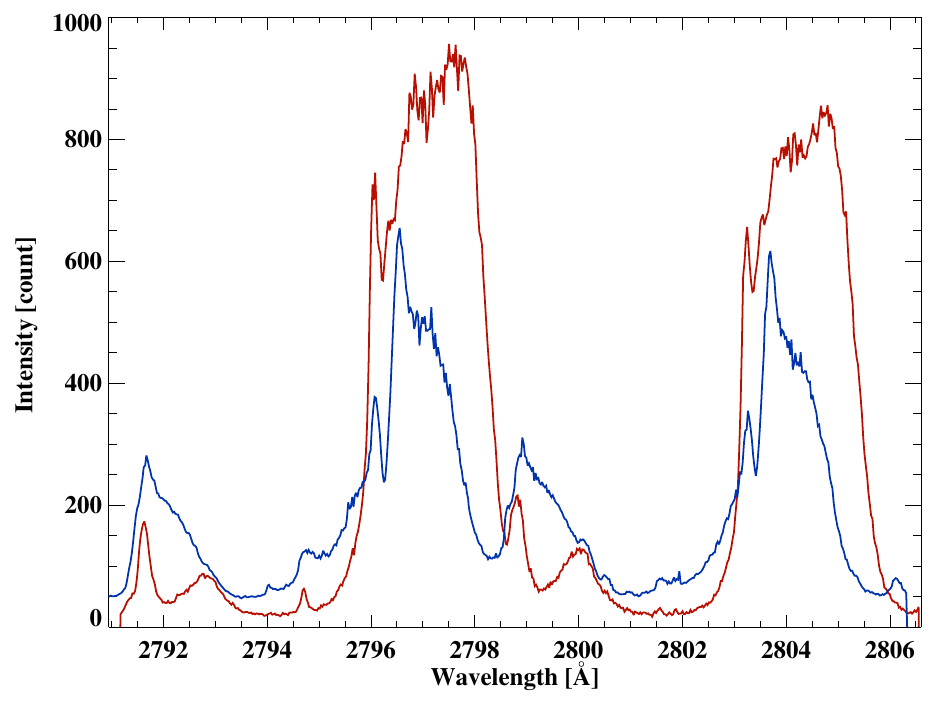}}
}
}
\caption{\textsl{\ion{Mg}{2} line profiles in 2014-March-29th flare (blue) at 17:46:13 UT on pixel 436 and 2022-March-30th flare (red) at 17:35:19 UT on pixel 51. The intensity of the line profile of 2014-March-29th flare is reduced by a factor of 10.}}
\label{Fig:comparison}
\end{figure}

Spectral profiles that resemble those presented here are not common in the literature, and those examples that have been reported were not as extreme as those presented here, nor did they compare the \ion{Si}{4}, \ion{C}{2} or \ion{O}{1} lines. \cite{Liu2015} analysed IRIS observations of the 2014-March-29th X-class solar flare, focusing on \ion{Mg}{2}, and found some profiles with large red wings, for example their Figure 6. However, the \ion{Mg}{2} line wings did not extend as far from line core as those presented here, as shown in Figure~\ref{Fig:comparison}. At t = 17:46:13, the \ion{Mg}{2} line profile showed a redshfit of about 75 km s$^{-1}$ (51 km s$^{-1}$ using spectral moment method). Then again at 17:52:28UT, the \ion{Mg}{2} looks more like our profiles with a red-shift about 70 km s$^{-1}$ (64 km s$^{-1}$ using spectral moment method), and would also be a re-brightening some few minutes after the initial phase. \cite{Lacatus2017} also presented profiles that have prominent, broad red wings, which were interpreted as turbulent coronal rain following a flare. In that instance, however, the red wing asymmetries persisted for tens of minutes, rather than the transient feature we identified here. Further, the \ion{Mg}{2} triplet lines were described as being rather weak in comparison to the resonance lines, and did not seem to exhibit red wing asymmetries.

Via the ratio of the \ion{Mg}{2} lines, \cite{Lacatus2017} determined that the triplet lines were optically thin, and that the resonance lines deviated from an optically thin ratio due to radiation anisotropies.  In our flare observations the \ion{Mg}{2} h \& k lines were optically thick, with a line ratio that decreases closer to unity in the red wing\footnote{The line ratio would increase towards a value of 2:1 for k:h in the event that the lines formed under optically thin conditions.}. Figure~\ref{Fig:wavetime_asymm} shows the temporal evolution of the \ion{Mg}{2} k:h line intensity ratio, as a function of wavelength. When the red wing asymmetry is present the ratio in the red wing decreases relative to the blue wing (compare panel (c) to the other panels). Note that in that figure, the features near $\pm1.9$~\AA, and other large jumps, are due to nearby spectral lines. That the broad wing is very optically thick could suggest a high density of material within the condensation. Interestingly, the ratio of the line integrated-intensities (over $-0.2 - 2$~\AA) of \ion{Mg}{2} 2799~\AA\ to  \ion{Mg}{2} 2791~\AA\ increases during the SOI, from a ratio $R_{\mathrm{sub}} \sim1.25$ to $R_{\mathrm{sub}} \sim1.6$. This suggests that the line is still in the optically thick regime \citep{Lacatus2017}, but there could be optically thin contributions from the condensation also. These ratios can act as observational constraints on any future modelling.

If we can interpret these Doppler shifts as being a consequence of downflowing plasma then what is the cause of this supersonic mass flow? Typical flare models that model energy deposition by nonthermal electron precipitation, thermal conduction, or Alfv\'enic waves, predict more modest speeds in the upper chromosphere. In those models, condensations through the transition region and chromosphere are on the order of 10-100~km~s$^{-1}$, which rapidly decelerate such that they are typically around a few 10s km~s$^{-1}$ within the mid-upper chromosphere itself \citep[e.g.][]{Fisher1985a, Fisher1985b, Fisher1985c, Fisher1989}. See also these loop modelling reviews: \cite{2022FrASS...960856K} \& \cite{2023FrASS...960862K}. In this flare, the NUV continuum, which is assumed to be part of the Balmer continuum and together with the optical continuum represents white light flare (WLF) emission, exhibits an initial impulsive increase co-temporal with the initial passage of the flare ribbon over the source of interest, which reduces in intensity rapidly (within 60~s). Shortly before the second brightening of our source of interest, there is another enhancement of the NUV continuum, this time weaker, broader and with a more gradual temporal evolution, that persists for many minutes. The cause of WLF enhancements is still debated, but an impulsive response is known to be associated with hard X-rays, and thus presumably particle precipitation. That the second brightening is weaker and more gradual places constraints on the chromospheric conditions into which energy is deposited and on the energy transport mechanisms at play, though further comment and study is beyond the focus of this paper which is the excessive red-wing asymmetry of \ion{Mg}{2}, \ion{C}{2}, and \ion{Si}{4}.

\cite{RubiodaCosta2017} suggested very fast downflows as a means to explain very broad and asymmetric \ion{Mg}{2} spectra from the 2014-March-29th X class flare that were similar in appearance to the profiles presented by us here, though not as extreme and without the flattened line wings. By manually varying the velocity in a flare atmosphere to introduce a large velocity (200~km~s$^{-1}$) in the upper chromosphere they were able to synthesise \ion{Mg}{2} emission reminiscent of the profiles from 2014-March-29th flare\footnote{It should be noted, though, that \cite{RubiodaCosta2017} did not update the gas density in their model when introducing this velocity gradient. As found in radiation hydrodynamics simulations of flares, such large condensations would accrue significant mass, increasing the optical depth of \ion{Mg}{2}, and likely reducing the impact of velocity smearing.}. In our scenario this velocity structure would need to be even more extreme, extending to the lower transition region where \ion{Si}{4} and \ion{C}{2} form also. However, it is not clear how to arrive such a velocity profile within the transition region and upper chromosphere naturally (i.e. self-consistently in a flare-driven simulation), how to maintain it long enough, and how to quench it before the lower-mid chromosphere is affected.

Although not common, some flare numerical modelling studies have produced such significant downflow speeds. \citet{Ashfield2021} modeled condensation speeds up to $\sim300$~km s$^{-1}$ by considering heating of the chromosphere via thermal conduction, with energy supplied for the duration of their simulation. They noted that their peak downflow speeds were excessive compared to most observations (recall that most observations do not seem to suggest downflows much greater than 10-100 of km~s$^{-1}$), particularly given the rather weak energy fluxes injected. The authors speculated that the pre-structure of their chromosphere (isothermal and fully ionised) was not very realistic and could act to reduce the flow speed. Of note here is that our inferred large flow speeds occur in a region that has been previously heated a short time prior, so perhaps the ionisation state of the chromosphere is more comparable to the \citet{Ashfield2021} experiments. We do note that it takes some time for bright EUV loops to appear near the SOI, which calls into question the scenario of energy transport via conduction. Of course, it might be that the emission measure of those loops is rather low, but then the question of what suppressed evaporation to fill the loops needs to be addressed. The work of \cite{Ashfield2021} was built upon foundations made by \cite{Fisher1989}, who determined analytical relationships between lifetimes and magnitude of chromospheric condensations and the energy flux delivered to the chromosphere during flares. \cite{Fisher1989} found that the peak downflow velocity, $v_{p}$, was related to the pre-injection chromospheric mass density, $\rho_{\mathrm{chrom}}$: $v_{p} \propto \rho_{\mathrm{chrom}}^{-1/3}$. Thus one means to produce an exceptionally large downflow is to inject flare energy into a very low density region of the chromosphere. In such a scenario, there could be an underdense region of the upper chromosphere/lower transition region that allows the condensation to reach large velocities before picking up mass and decelerating.

Condensation velocities up to 150 km s$^{-1}$ \citep{Zhu2019} or higher  \citep{2018ApJ...852...61K} were produced in radiation hydrodynamic models driven by very large non-thermal electron beam fluxes ($5\times10^{11}$ and $1\times10^{13}$~erg~s$^{-1}$~cm$^{-2}$, respectively). In the former scenario, the downflow rapidly reduces in speed once it hits the upper chromosphere and accrues mass, dropping to a few tens of km~s$^{-1}$ within  2-3 seconds. The synthetic \ion{Mg}{2} spectra from \citep{Zhu2019} are not consistent with the observations presented here. Initially their synthetic profiles are entirely redshifted, moving rapidly towards the rest component. Multiple line components are only present for a very short time. Our observations, however, exhibit a mostly stationary component throughout, alongside a strong enhancement to the red wing. These persist for multiple 2.5s exposures, up to 60~s in duration. So, while these high-flux particle beam experiments can produce the required velocities, those downflows are thus far too transient to arrive at the velocity stratification seemingly demanded by our observations.

Finally we note some similarities and differences between our observations and those of \cite{Ichimoto1984}. Those authors found localised ($\sim1$\arcsec) H$\alpha$ flare sources which exhibited strong red wing asymmetries. Typical inferred Doppler motions were smaller than those found in this flare (around $40-100$~km~s$^{-1}$), though their Figure 6c does indicate potentially higher values up to 140~km~s$^{-1}$. Those localised fast flows were short-lived, decreasing rapidly to a few 10s km~s$^{-1}$ within 30s. Indeed, the H$\alpha$ profiles presented by \cite{Ichimoto1984} at times appear reminiscent of the profiles shapes discussed by us (e.g. their Figure 4a). Those authors attributed their observations to chromospheric condensations. If this is also true of our observations, that appear to be more extreme examples and which pose a problem to current electron beam or conduction driven flare simulations, then we must endeavour to understand the circumstances that lead to the development of such a strong velocity.
\section{Summary}

In this study, we study a transient emission with unusually strong enhancement in the red-wing, that could be interpreted as downward velocities about 40 - 160~km~s$^{-1}$. To our knowledge, such high supersonic downflows have not been reported commonly in flare footpoints. The emission source was located behind one of the major flare ribbon moving mainly southward, in a region that had previously been heated by the flare, such that the ribbon appears to extend back on itself. The spectral features, and images of the emission enhancement, suggest a lifetime of approximately $41 \pm 15$~s. Since the profiles were flattened along the red wings, it is reasonable to assume that the density of the condensation was large enough to raise the $\tau_{\lambda} = 1$ height throughout the entire wing such that the wing formed entirely within the condensation.

A quantitative analysis was performed, using spectral moments and Gaussian fitting, to attempt to characterize the typical Doppler motions involved, but we stress that the interpretation of Doppler motions from optically thick lines to ascribe an actual atmospheric bulk motion is complicated. That the \ion{C}{2} and \ion{Si}{4} lines (the latter of which may be optically thin) give metrics consistent with \ion{Mg}{2} does give confidence in our interpretation, and even if the absolute values of the downflow can not be stated without the caveats of dealing with optically thick lines, these red wings are certainly more extreme that is typical in flare observations

Extreme excursions into the red wings had an a rapid onset time, and seemed confined to the transition region and upper chromosphere, evidenced by the lack of contemporaneous shifted emission of the \ion{O}{1} line. The increase of opacity in the \ion{Mg}{2} resonance line wings could be evidence of increased density. Given (1) the similarity in appearance to observations of chromospheric condensations \citep[e.g.][who showed separated components merging with the stationary component]{Graham2020}, (2) that the \ion{Mg}{2} subordinate lines are strongly in emission and also exhibit asymmetries \citep[unlike the weak profiles from the coronal rain observations of ][]{Lacatus2017}, and (3) that the appearance of a prominent red wing asymmetry in the \ion{Mg}{2} subordinate lines take somewhat longer to appear, we speculate that this observation is an extreme example of a dense chromospheric condensation. This condensation subsequently propagates through the transition region and chromosphere but is damped in the mid-upper chromosphere. Such a supersonic condensation would likely lead to a shock and subsequent heating, but it cools very rapidly and the lower atmosphere is seemingly not strongly affected. Its speed is at odds with both typical observations and also typical modelling of chromospheric condensations. Though models do suggest such large velocities can be produced, they are thus far too transient and subside to $10-100$~km~s$^{-1}$ upon accruing mass in the upper chromosphere. Modelling of the flaring chromosphere should endeavour to explore the conditions that leads to these supersonic flows, with the fact that this source occurred in a previously heated atmosphere is perhaps important. Based on theoretical analysis of condensations \citep[][]{Fisher1989,Ashfield2021}, this condensation, although seemingly dense when emitting in \ion{Mg}{2}, may have been generated following energy injection into an initially under-dense region.\\

\textsc{Acknowledgments:}

We thank the anonymous referee for their useful comments, which improved the clarity of this manuscript. This work is supported by NSF under grants AGS 1821294, 1954737, 1936361, 2228996 2309939 and AST 2108235, and by the NASA under grants 80NSSC17K0016, 80NSSC19K0257, 80NSSC19K0859, 80NSSC21K1671, 80NSSC20K0716, and 80NSSC21K0003. VP acknowledges financial support from the NASA ROSES Heliophysics Guest Investigator program (Grant\# NASA 80NSSC20K0716). GSK acknowledges financial support from NASA's Early Career Investigator Program (Grant\# NASA 80NSSC21K0460). IRIS is a NASA Small Explorer mission developed and operated by LMSAL with mission operations executed at NASA Ames Research center and major contributions to downlink communications funded by the Norwegian Space Center (NSC,Norway) through an ESA PRODEX contract.

\appendix

\section{Source of interest evolution in 1330~\AA}\label{sec:sourceevol}

\begin{figure*}
	\centering
	\vbox{
	\hbox{
	\subfloat{\includegraphics[width = 0.315\textwidth, clip = true, trim = 0.cm 0.cm 0.cm 0.cm]{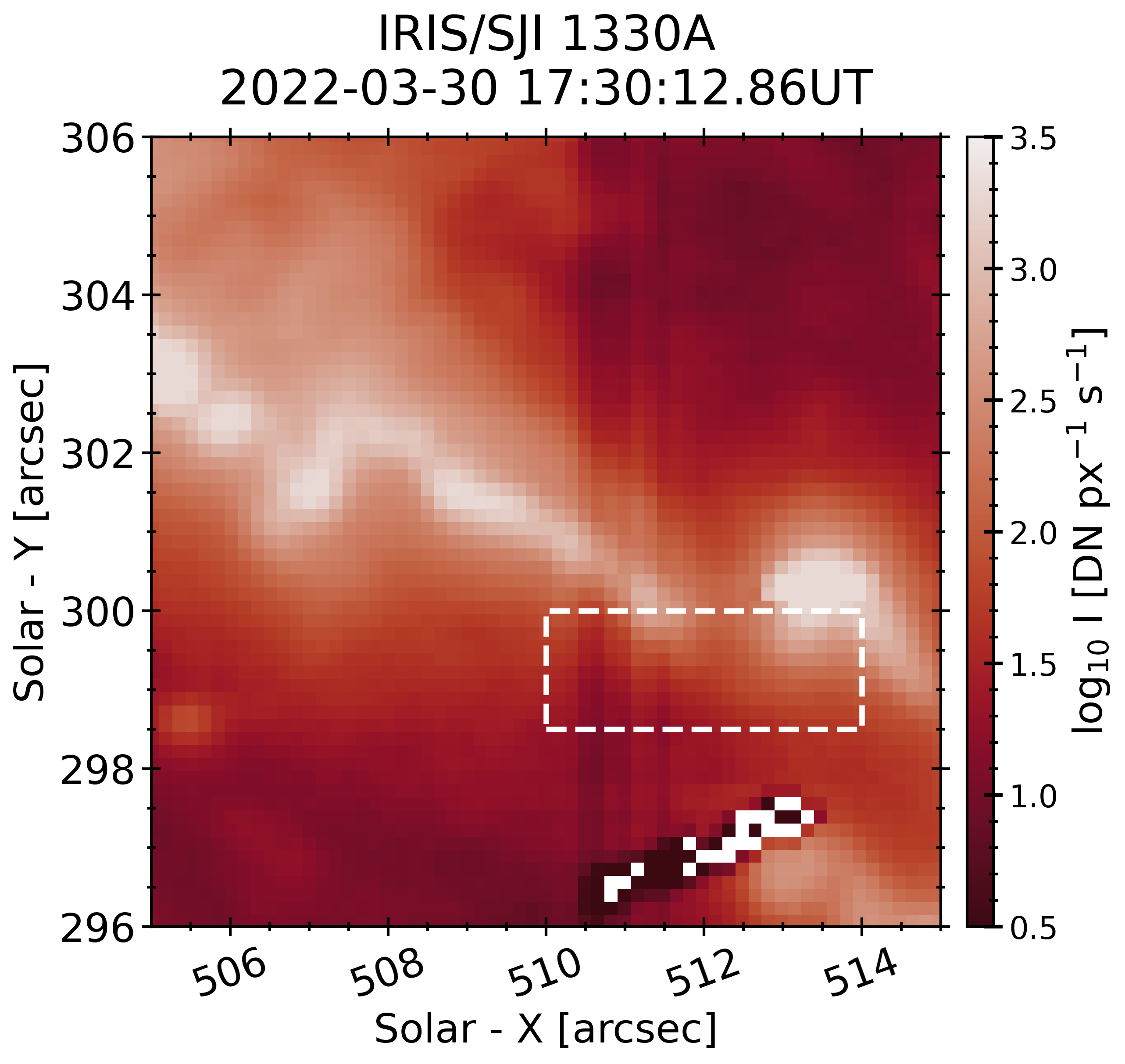}}
	\subfloat{\includegraphics[width = 0.315\textwidth, clip = true, trim = 0.cm 0.cm 0.cm 0.cm]{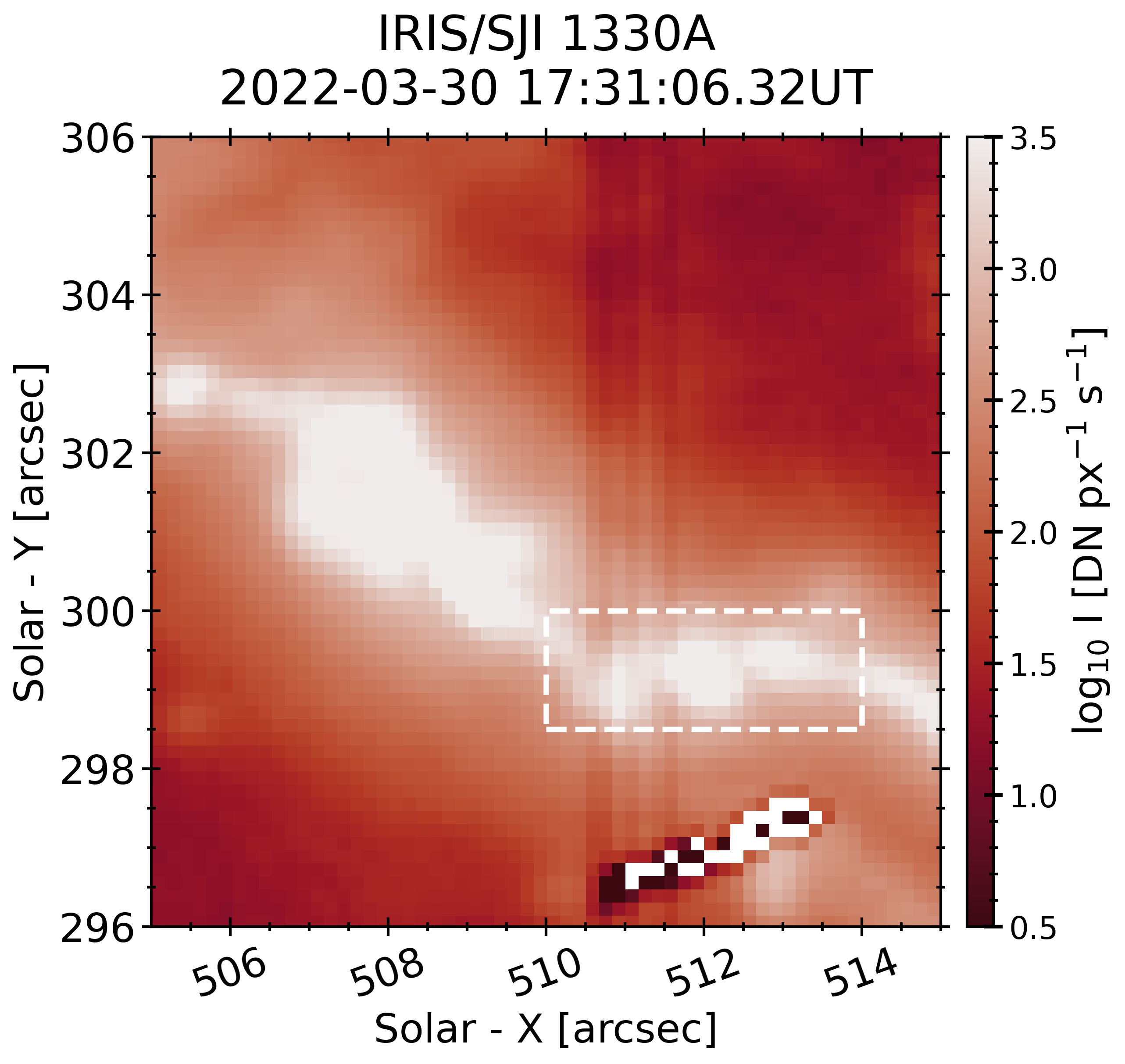}}	
    \subfloat{\includegraphics[width = 0.315\textwidth, clip = true, trim = 0.cm 0.cm 0.cm 0.cm]{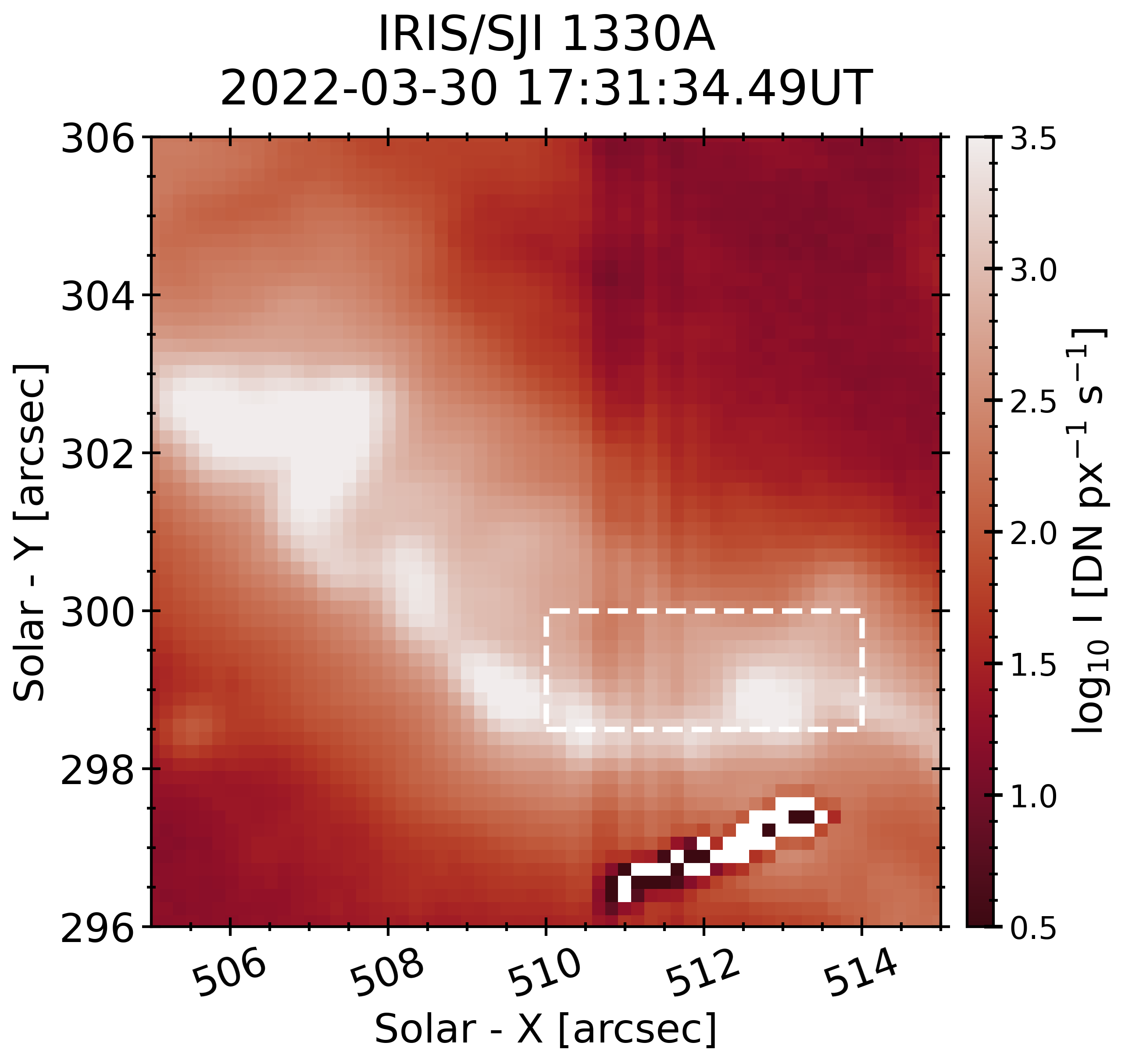}}	
        }
        }
        \vbox{
	\hbox{
 	\subfloat{\includegraphics[width = 0.315\textwidth, clip = true, trim = 0.cm 0.cm 0.cm 0.cm]{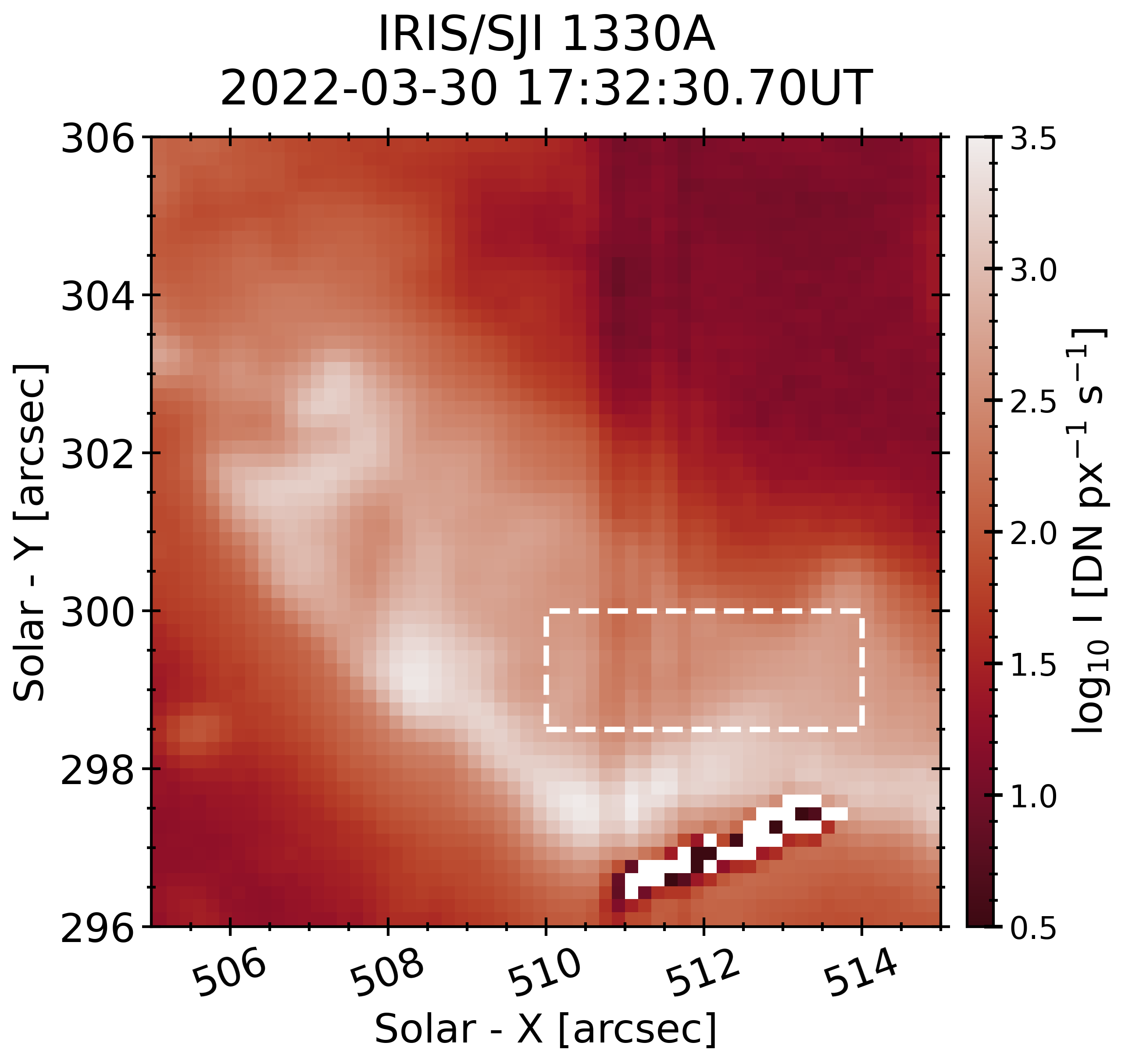}}
	\subfloat{\includegraphics[width = 0.315\textwidth, clip = true, trim = 0.cm 0.cm 0.cm 0.cm]{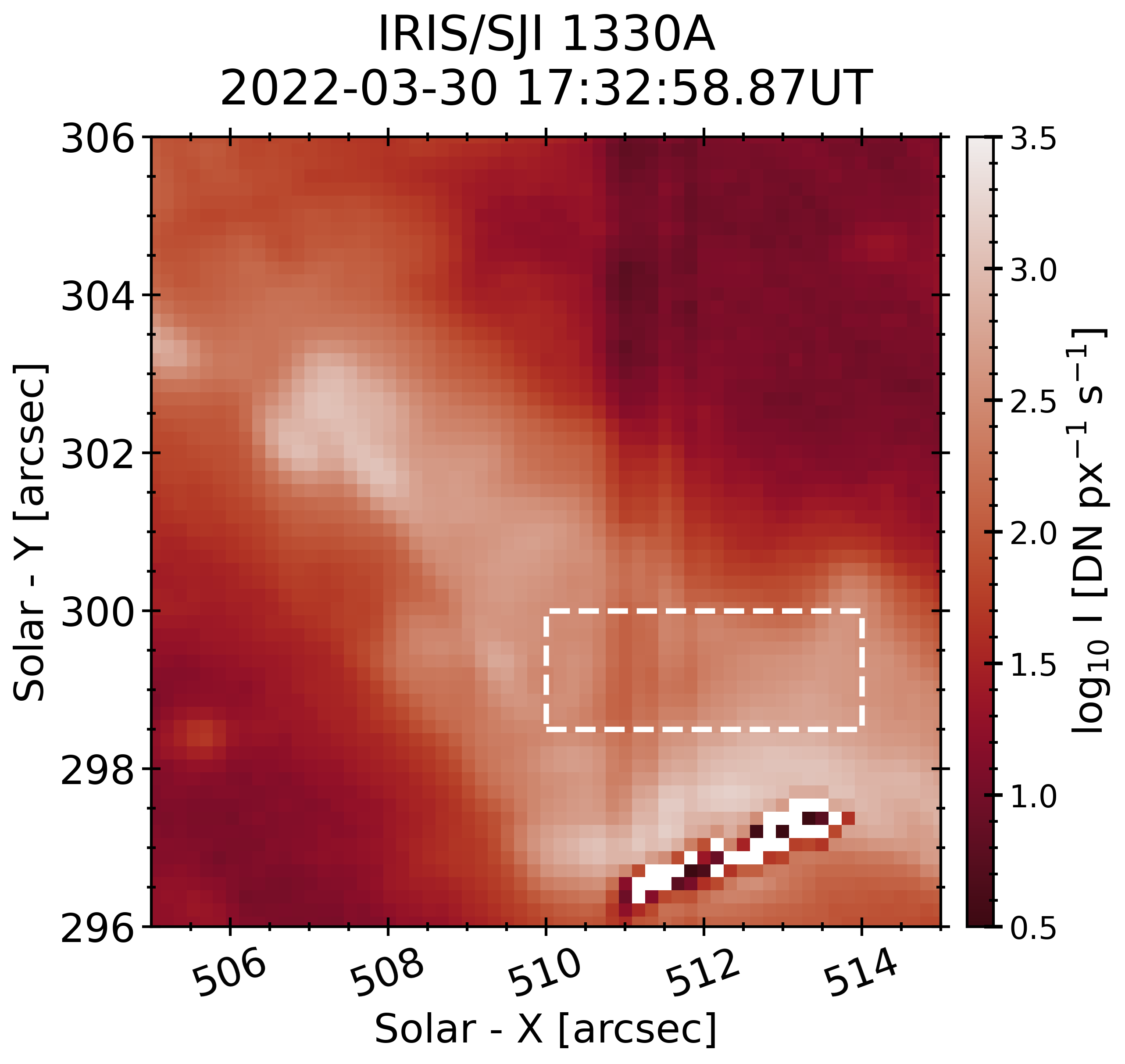}}	
    \subfloat{\includegraphics[width = 0.315\textwidth, clip = true, trim = 0.cm 0.cm 0.cm 0.cm]{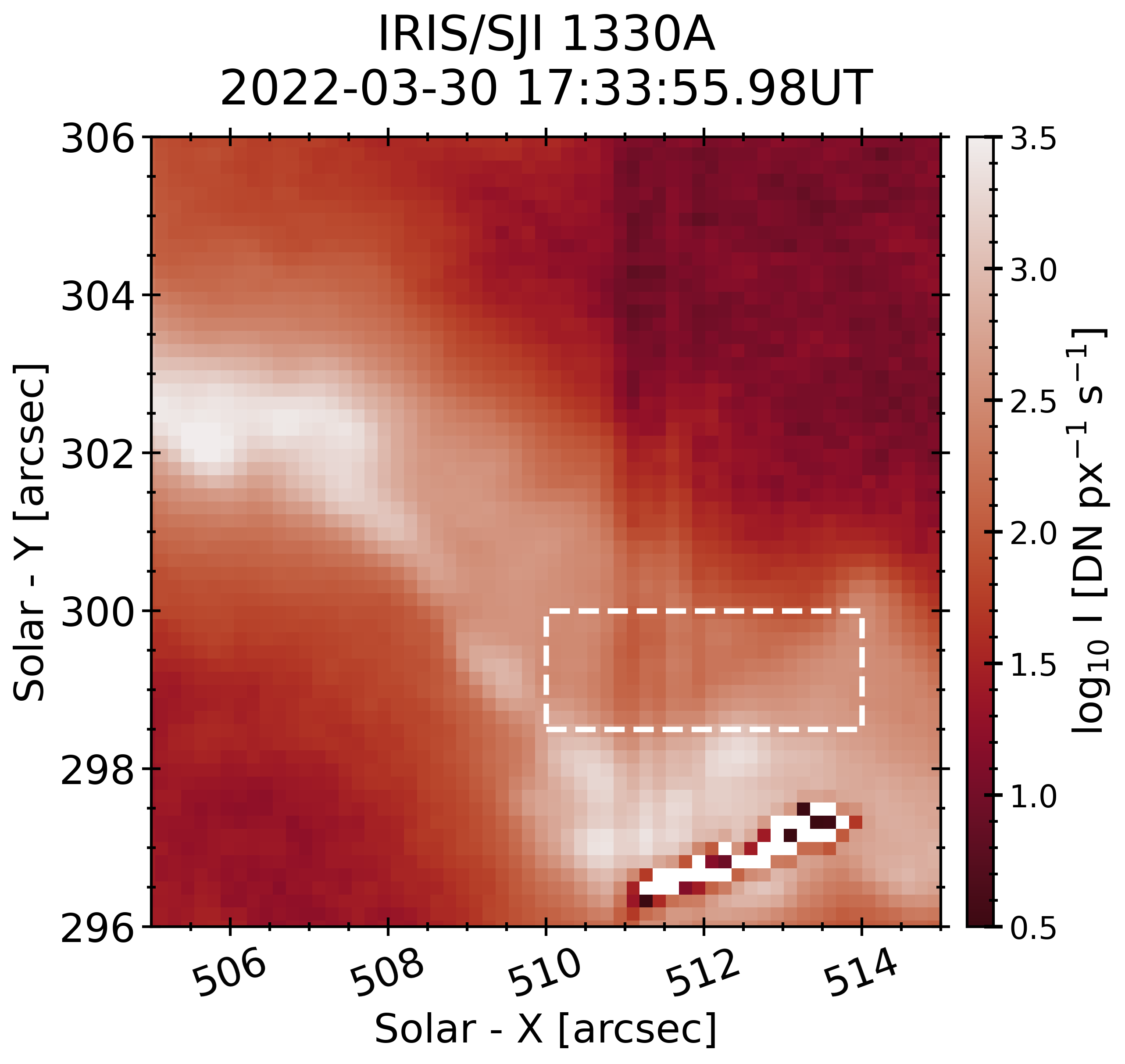}}	
        }
        }
        \vbox{
	\hbox{
 	\subfloat{\includegraphics[width = 0.315\textwidth, clip = true, trim = 0.cm 0.cm 0.cm 0.cm]{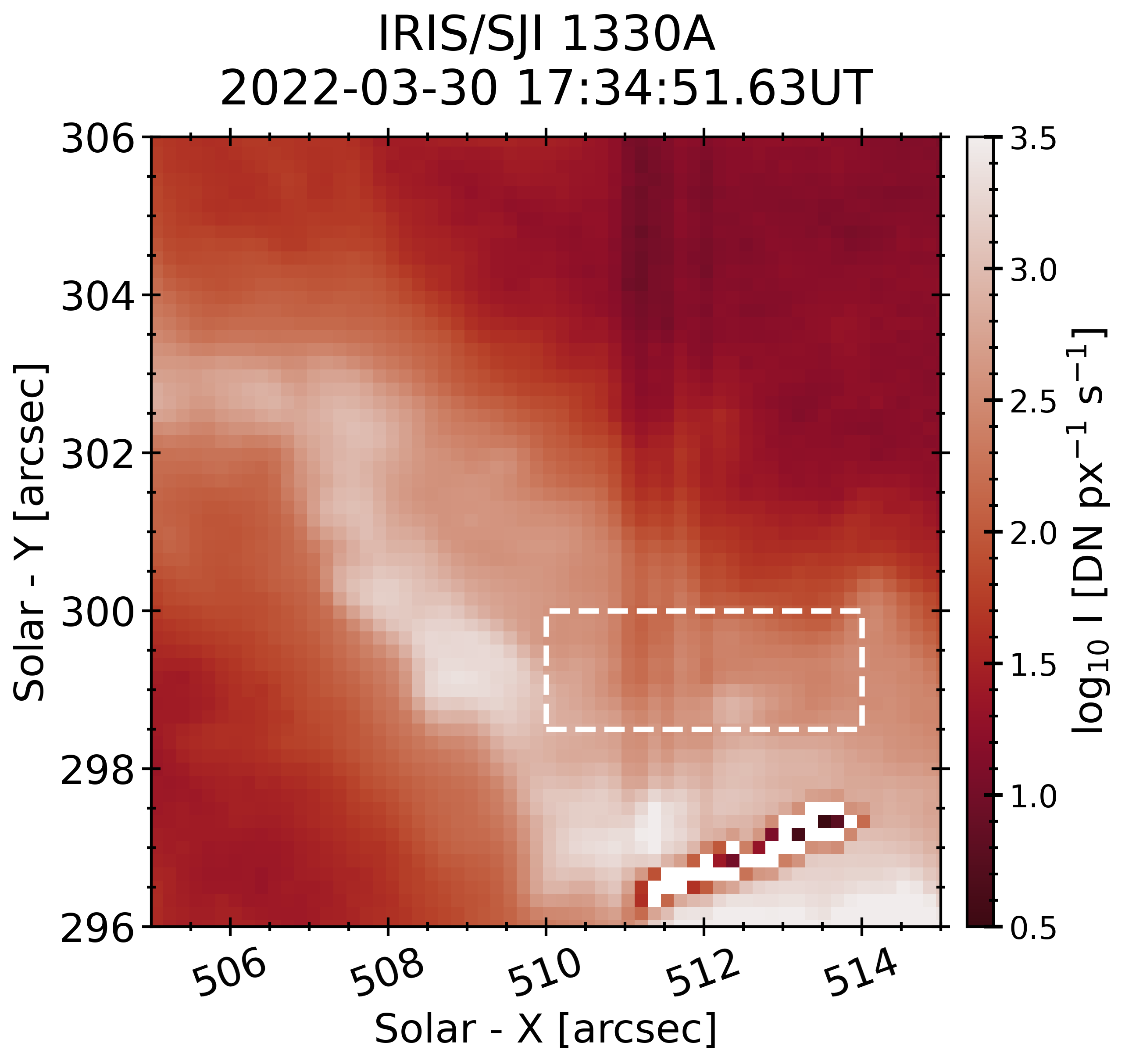}}
	\subfloat{\includegraphics[width = 0.315\textwidth, clip = true, trim = 0.cm 0.cm 0.cm 0.cm]{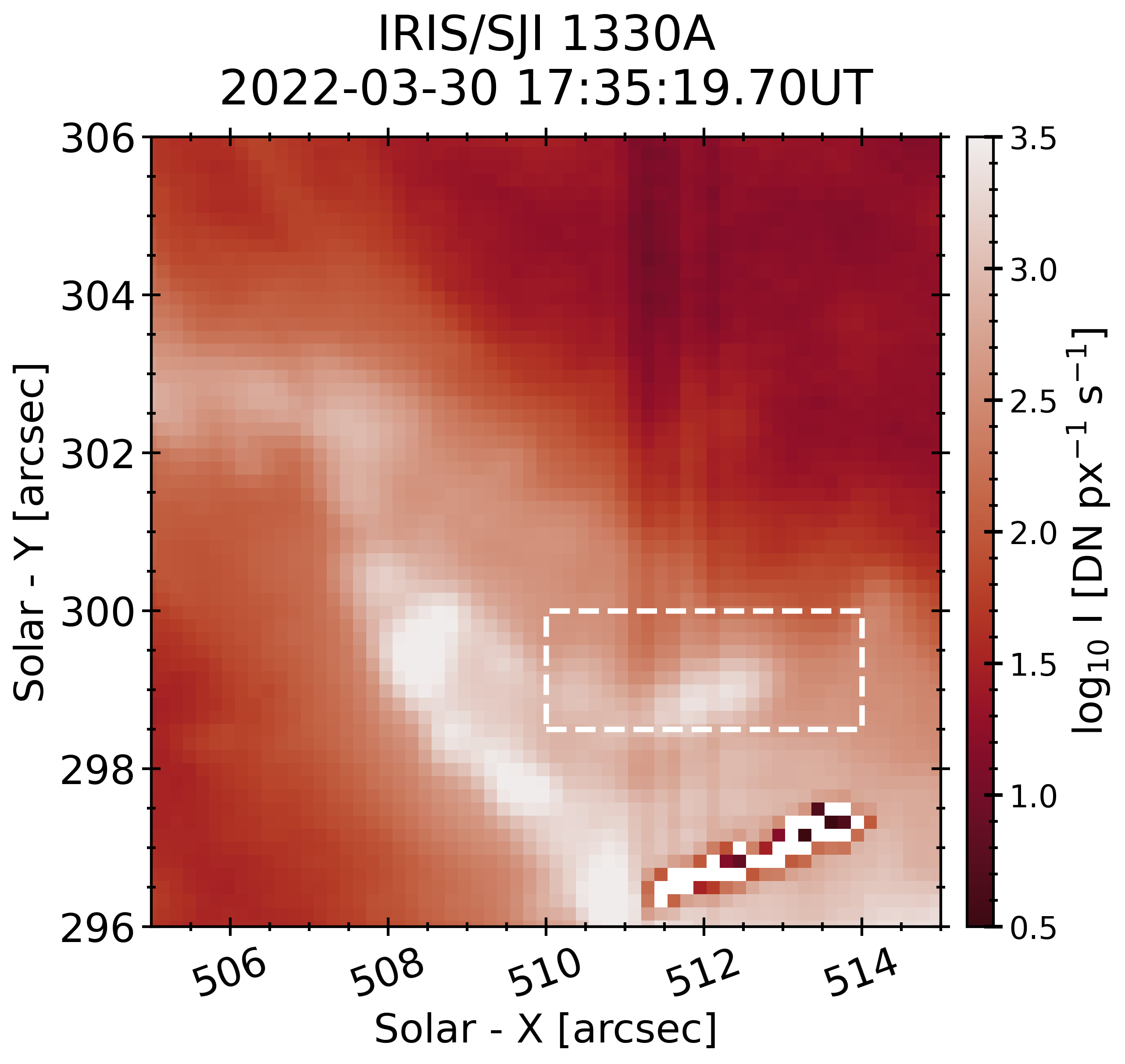}}	
    \subfloat{\includegraphics[width = 0.315\textwidth, clip = true, trim = 0.cm 0.cm 0.cm 0.cm]{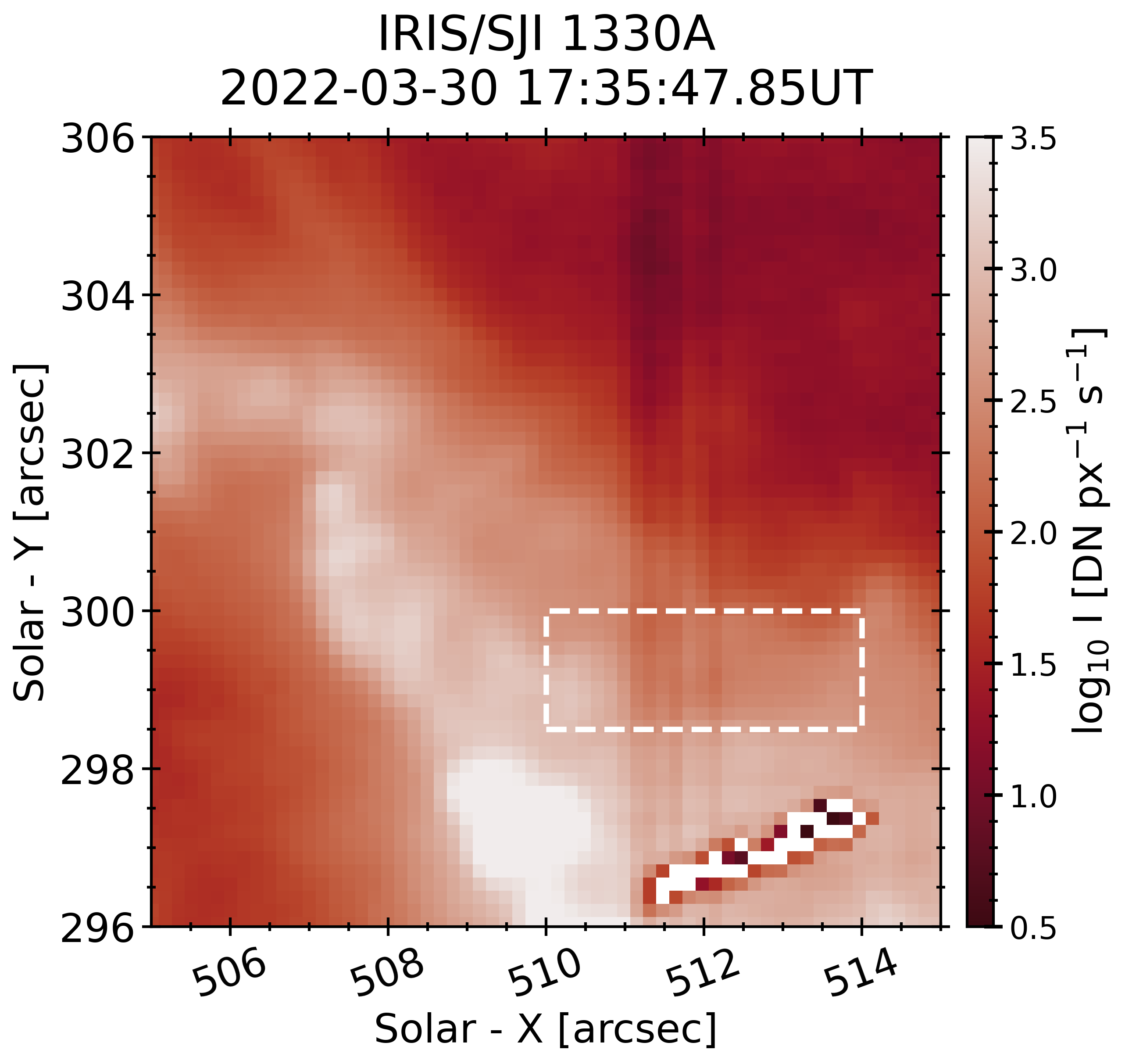}}	
        }
        }
    \caption{\textsl{Same as Figure~\ref{Fig:SJIoverview_2796} but for the IRIS 1330~\AA\ SJI passband. The  distorted feature in the bottom of the image, extending from x = 510\arcsec-514\arcsec\ is the result of dust on the detector, which appears bright here since the image is shown on a logarithmic scale}}
	\label{Fig:SJIoverview_1330}
\end{figure*}

The temporal evolution of the initial brightening and re-appearance of an enhanced feature near the source of interest as observed by the 1330~\AA\ SJI filter is shown in Figure~\ref{Fig:SJIoverview_1330}.

\section{sec:Spectra during the initial ribbon brightening}\label{sec:prespectra}

\begin{figure*}
	\centering
	\vbox{
	\hbox{
	\hspace{0.75in}
	\subfloat{\includegraphics[width = 0.75\textwidth, clip = true, trim = 0.cm 0.cm 0.cm 0.cm]{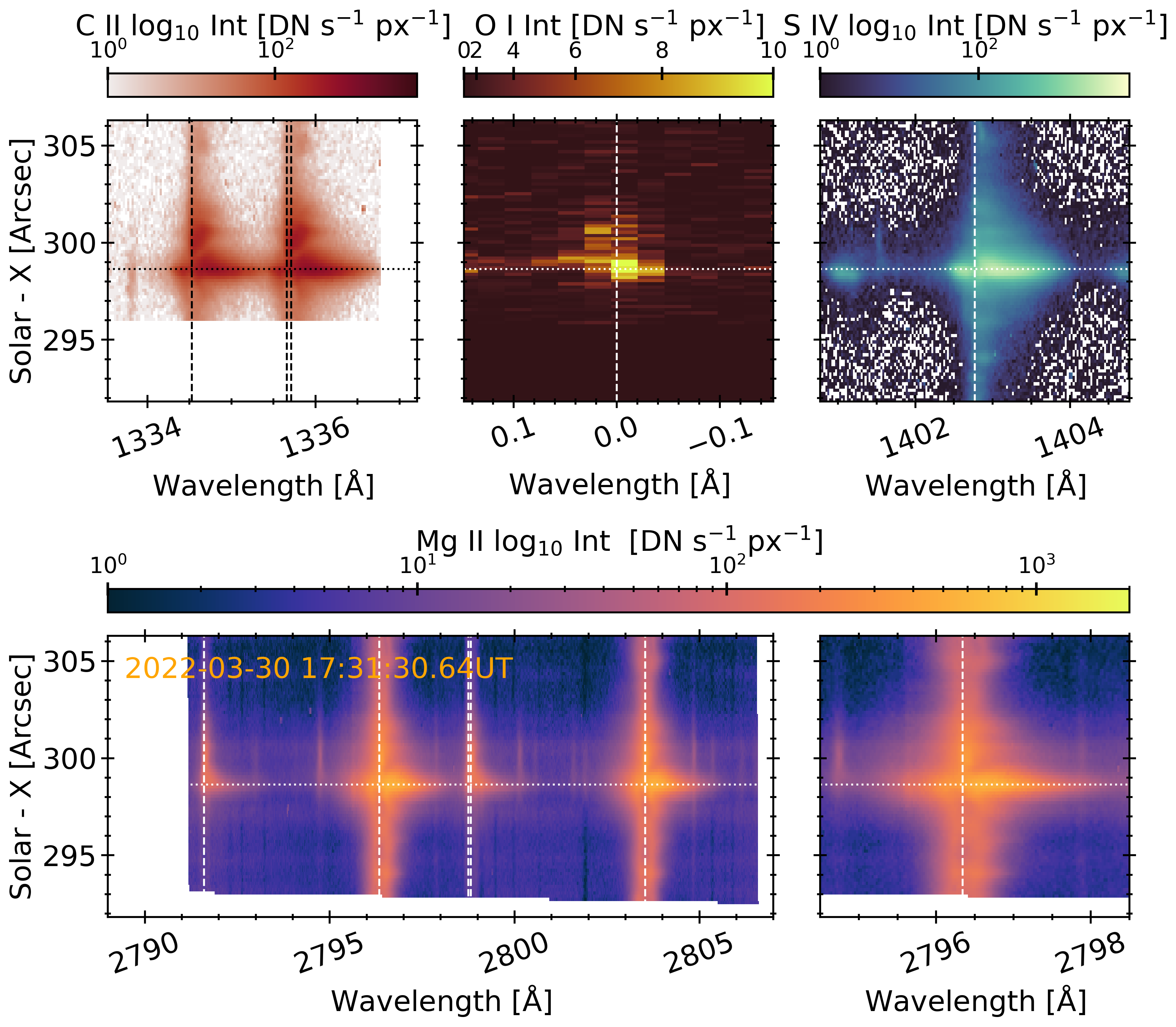}}
        }
        }
    \caption{\textsl{Same as Figure~\ref{Fig:SGoverview_SOI}, but during the initial brightening of the flare ribbon that passes over the source of interest. The dotted horizontal line indicates the location of the source of interest.}}
	\label{Fig:SGoverview_pre}
\end{figure*}

Before the brightening in the source of interest at around 1735~UT, the southern flare ribbon swept past at around 1731~UT. The spectra during that time were less extreme than during the SOI, and are shown in Figure~\ref{Fig:SGoverview_pre}.
\end{document}